\documentclass{article} 
\usepackage{iclr2023_conference,times}

\usepackage{amsmath,amsfonts,bm}

\def\eqref#1{equation~\ref{#1}}

\def\1{\bm{1}}

\def\vb{{\bm{b}}}

\def\vy{{\bm{y}}}

\def\mC{{\bm{C}}}

\def\mI{{\bm{I}}}

\def\mM{{\bm{M}}}

\def\mP{{\bm{P}}}

\def\mT{{\bm{T}}}

\def\mW{{\bm{W}}}
\def\mX{{\bm{X}}}

\DeclareMathAlphabet{\mathsfit}{\encodingdefault}{\sfdefault}{m}{sl}
\SetMathAlphabet{\mathsfit}{bold}{\encodingdefault}{\sfdefault}{bx}{n}
\newcommand{\tens}[1]{\bm{\mathsfit{#1}}}

\def\tR{{\tens{R}}}

\def\tT{{\tens{T}}}

\usepackage{hyperref}
\usepackage{url}
\usepackage{microtype}
\usepackage{graphicx}
\usepackage{caption}
\usepackage{booktabs}  
\usepackage{bm}  
\usepackage{enumitem}  
\usepackage{adjustbox}  
\usepackage{siunitx}
\usepackage{bbm}  
\usepackage[section]{placeins}  

\title{\mindmusic: Reconstructing Music from\\Human Brain Activity}

\author{
\hspace{4.4cm}\textbf{Timo I. Denk$^{*\text{ }1}$ ~~~ Yu Takagi$^{*\text{ }2,3}$} \vspace{0.1cm}\\
\hspace{2.2cm}\textbf{Takuya Matsuyama$^{2}$ ~~~ Andrea Agostinelli$^{1}$ ~~~ Tomoya Nakai$^{4}$}\vspace{0.1cm}\\
\hspace{3.88cm}\textbf{Christian Frank$^{1}$ ~~~ Shinji Nishimoto$^{2,3}$}\vspace{0.1cm}\\
\hspace{1.3cm}{$^1$ Google ~~~ $^2$ Osaka University, Japan ~~~ $^3$ NICT, Japan ~~~ $^4$ Araya Inc., Japan}}

\newcommand\blfootnote[1]{%
  \begingroup
  \renewcommand\thefootnote{}\footnote{#1}%
  \addtocounter{footnote}{-1}%
  \endgroup
}

\newcommand{\mindmusic}{Brain2Music}
\newcommand{\mulan}{MuLan}
\newcommand{\mulantext}{MuLan$^\text{text}$}
\newcommand{\mulanmusic}{MuLan$^\text{music}$}
\newcommand{\musiclm}{MusicLM}
\newcommand{\wvbert}{w2v-BERT}
\newcommand{\soundstream}{SoundStream}

\newcommand{\baseurl}{https://google-research.github.io/seanet/brain2music/}
\newcommand{\displayurl}{google-research.github.io/seanet/brain2music}
\newcommand{\website}[1]{\href{\baseurl#1}{\displayurl#1}}

\iclrfinalcopy 
\begin{document}
\maketitle
\begin{abstract}
The process of reconstructing experiences from human brain activity offers a unique lens into how the brain interprets and represents the world.
In this paper, we introduce a method for reconstructing music from brain activity, captured using functional magnetic resonance imaging (fMRI).
Our approach uses either music retrieval or the {\musiclm} music generation model conditioned on embeddings derived from fMRI data.
The generated music resembles the musical stimuli that human subjects experienced, with respect to semantic properties like genre, instrumentation, and mood.
We investigate the relationship between different components of {\musiclm} and brain activity through a voxel-wise encoding modeling analysis.
Furthermore, we discuss which brain regions represent information derived from purely textual descriptions of music stimuli.
We provide supplementary material including examples of the reconstructed music at \website{}
\blfootnote{$^*$ Equal contribution; correspondence to timodenk@google.com and takagi.yuu.fbs@osaka-u.ac.jp}
\end{abstract}

\begin{figure}[h]
\begin{center}
\includegraphics[trim=.7cm 8.5cm 3cm 2cm, clip, width=0.9\textwidth]{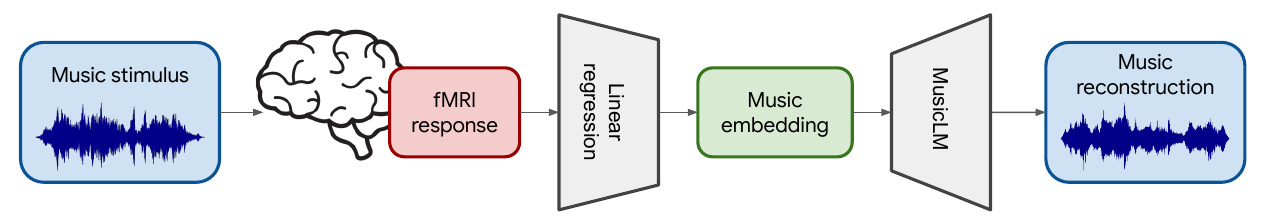}
\end{center}
\caption{An overview of our {\mindmusic} pipeline: High-dimensional fMRI responses are condensed into the semantic, 128-dimensional music embedding space of {\mulan} \citep{mulan}. Subsequently, the music generation model, {\musiclm} \citep{musiclm}, is conditioned to generate the music reconstruction, resembling the original stimulus. As an alternative we consider retrieving music from a large database, instead of generating it.}
\label{fig:decoding-pipeline}
\end{figure}

\section{Introduction}
\label{sec:introduction}

Music holds universal significance, acting as a medium for expression and communication across diverse cultures. The representation of music within our brains has been a significant topic of interest in neuroscience. Previous studies have examined human brain activity, as captured by functional magnetic resonance imaging (fMRI), while participants listened to music. These studies discovered musical feature representations in the brain, such as rhythms \citep{ALLURI20123677}, timbres \citep{TOIVIAINEN2014170,ALLEN201860}, emotions \citep{koelsch2006investigating}, and musical genres \citep{casey2017music,nakai2021music}. This body of research provides valuable insights into how music-related features -- both objective and subjective, acoustic and semantic -- are represented within the brain, illuminating the complexity of our experiences with music.

With the advent of text-to-music models, conditional generation of high-fidelity music has become feasible \citep{riffusion,musiclm,noise2music,tiktokmusicgen,fbmusicgen}. This exciting development bridges the gap between our linguistic understanding of music and the actual creation of musical compositions. Thus, new questions arise: How do the text and music embeddings used by these music generation models correspond to representations in the human brain? Furthermore, if a correspondence exists, is it possible to generate music directly from brain activity?

In this paper, we explore the feasibility of reconstructing music from brain activity scans with {\musiclm}~\citep{musiclm}.
We also compare some of the internal representations of {\musiclm}, such as the high-level semantic embedding space of {\mulan}~\citep{mulan} and the low-level, averaged embeddings from {\wvbert}, to activity in different brain regions, providing novel insights.

In detail, we make the following contributions:

\begin{itemize}
    \item We reconstruct music from fMRI scans by predicting high-level, semantically-structured music embeddings and using a deep neural network to generate music from those features. Our evaluation indicates that the reconstructed music semantically resembles the original music stimulus.
    \item We find that different components of our music generation model are predictive of activity in the human auditory cortex. The distinction between low- and high-level audio representations within the auditory cortex is less pronounced than that observed in the visual cortex for visual stimuli.
    \item We offer novel insights suggesting that within the auditory cortex there is significant overlap in the voxels that are predictable from (a)~purely textual descriptions of music and (b)~the music itself.
\end{itemize}

\section{Related Work}
\label{sec:related-work}

\subsection{Music Generation Models}

Generating high-fidelity music has been challenging due to the need to produce music with both high-quality audio and long-term consistency. An initial approach was introduced by Jukebox \citep{jukebox}, which proposes a hierarchical structure at different time resolutions, modeled autoregressively with Transformer-based models. While Jukebox generates music with high temporal coherence, it contains perceptible artifacts. PerceiverAR~\citep{perceiverar} makes use of {\soundstream}, a neural audio codec which compresses audio at low bitrates, while maintaining a high-quality reconstruction~\citep{soundstream}. PerceiverAR models a sequence of discrete SoundStream embeddings autoregressively, generating high quality audio but lacking in temporal coherence.

Recently, auto-regressive and diffusion-based models have significantly advanced the quality of synthesis in both music and broader audio generation. AudioLM~\citep{audiolm} suggests autoregressively modeling a hierarchical tokenization scheme composed of both \textit{semantic} and \textit{acoustic} discrete audio representations. MusicLM~\citep{musiclm} integrates the AudioLM framework with the joint music/text embedding model {\mulan} \citep{mulan}, enabling the generation of high-fidelity music conditioned on detailed text descriptions.

Other Transformer-based methods encompass \cite{singsong}, \cite{tiktokmusicgen}, and \cite{fbmusicgen}, with the last of these leveraging the EnCodec audio codec \citep{encodec}. Additionally, diffusion model-based strategies for music generation are used by Riffusion \citep{riffusion}, with the most recent advancements proposed in \citep{noise2music, liu2023audioldm, ghosal2023text, lam2023efficient}.

Within the framework of our {\mindmusic} pipeline, we employ {\musiclm} and its constituent components. However, our methodology is fundamentally adaptable to any music generator, provided the generator can accommodate conditioning on a dense embedding. When this project began, {\musiclm} was the only potent music generation model accessible to us.

\subsection{fMRI Audio Decoding and fMRI Encoding}

One of the key goals in neuroscience is to understand the representations that govern the relationship between brain activity and sensory and cognitive experiences. To this end, researchers construct encoding models to quantitatively describe which features of these experiences (e.g., color, motion, and phonemes) are encoded as brain activity. In contrast, they also build decoding models to infer the experienced content from a specific pattern of brain activity (for a review, see \cite{naselaris2011}).

Particularly in recent years, researchers have discovered correspondences between the internal representations of deep learning models and those of the brain across various sensory and cognitive modalities \citep{yamins2014performance, Kell2018}.
These findings have advanced our understanding of brain functions through (a)~the development of encoding models mediated by the representations \citep{gucclu2015deep}, (b)~interpretations of the representations based on their correspondence with brain functions \citep{Cross2021, image-reconstruction-takagi}, and (c)~reconstruction of experienced content (such as visual images) from brain activity  \citep{Shen2019, Chen_2023_CVPR, Takagi2023TechRep}.

More specifically, in the context of investigating auditory brain functions, researchers have developed encoding models using deep learning models that process auditory inputs \citep{Kell2018}, and conducted studies to reconstruct perceived sounds from brain activity \citep{Santoro2017, park2023sound}.
However, so far, these studies have largely targeted general sounds, including voices and natural sounds.
There are no instances of constructing encoding models using the internal representations of text-to-music generative models, or reconstructing musical experiences from brain activity with a focus on music and its unique features.

\section{Methods}
\label{sec:method}

\subsection{Music fMRI Dataset}
We pre-process the \textit{music genre neuroimaging dataset}\footnote{Download link: \href{https://openneuro.org/datasets/ds003720/versions/1.0.0}{openneuro.org/datasets/ds003720}} from \cite{nakaidataset} in a same manner as \cite{nakai2021music}. See Appendix~\ref{sec:appendix-fMRI-preprocess} for details of the preprocessing protocol.
The dataset contains music stimuli from 10~genres (blues, classical, country, disco, hip-hop, jazz, metal, pop, reggae, and rock) which were sampled randomly from the (music-only) dataset GTZAN \citep{gtzan}.
A total of 54 music pieces (30s, 22.050kHz) were selected from each genre, providing 540~music pieces.
A 15s~music clip was selected at random from each music piece. For each clip, 2s of fade-in and fade-out were applied and the overall intensity was normalized. The dataset contains 480 examples for training and 60 for reporting the final results.

\paragraph{Data collection details.} During scanning, five participants were asked to focus on a fixation cross at the center of the screen and to listen to the music clips through MRI-compatible insert earphones (Model S14, Sensimetrics). Every subject heard the same music clips. This headphone model can attenuate scanner noise and has been widely used in previous MRI studies with auditory stimuli \citep{norman2015distinct}.
Scanning was performed using a 3.0T~MRI scanner (TIM~Trio; Siemens, Erlangen, Germany) equipped with a 32-channel head coil.
For functional scanning, we scanned 68~interleaved axial slices with a thickness of 2.0mm without a gap using a T2$^*$-weighted gradient echo multi-band echo-planar imaging (MB-EPI) sequence \citep{moeller2010multiband} (repetition time~(TR, aka. sampling interval)~=~1,500ms, echo time~(TE)~=~30ms, flip angle~(FA)~=~$62^\circ$, field of view~(FOV)~=~$192\times192\text{mm}^2$, voxel size~=~$2\times2\times2\text{mm}^3$, multi-band factor~=~4).
A total of 410~volumes were obtained for each run.

\paragraph{Text captions.} In this work, we augment the original dataset \citep{nakaidataset} by introducing English text captions which we have made publicly available\footnote{\href{https://www.kaggle.com/datasets/nishimotolab/music-caption-brain2music}{kaggle.com/datasets/nishimotolab/music-caption-brain2music}}.
These captions, averaging approximately 46 words or 280 characters in length, typically describe the musical pieces in terms of genre, instrumentation, rhythm, and mood. They often comprise fragmented or semi-complete sentences, with an average of about 4.5 sentences per caption.
The style of writing is subjective, reflecting not only the technical components of the music such as the instruments used, but also the emotional responses or atmospheres they might evoke in listeners. Captions were written in Japanese and translated using DeepL. Several exemplar captions and the instructions given to the raters are in Section~\ref{sec:appendix-caption-data}.
In this paper, the captions are used to study how purely semantic, text-derived embeddings relate to brain activity induced by the corresponding music stimuli.

\subsection{{\mulan} and {\musiclm}}
\label{sec:musiclm}

{\mulan} \citep{mulan} is a joint text/music embedding model consisting of two towers, one for text (\mulantext) and one for music (\mulanmusic).
The text tower is a BERT \citep{devlin2019bert} model pre-trained on a large text corpus. For the audio tower we use the ResNet-50 \citep{he2015deep} variant.
{\mulan}'s training objective is to minimize a contrastive loss between the 128-dimensional embeddings produced by each tower for an example pair of aligned music and text.
For example, the embedding of a rock song's waveform is supposed to be close to the embedding of the text \textit{rock music} and far from \textit{calming violin solo}.
In this paper, if we refer to a {\mulan} embedding we mean by default the embedding of the music tower.

\begin{figure}[t]
\begin{center}
\includegraphics[trim=3cm 4.7cm 5.5cm 2.7cm, clip, width=0.9\textwidth]{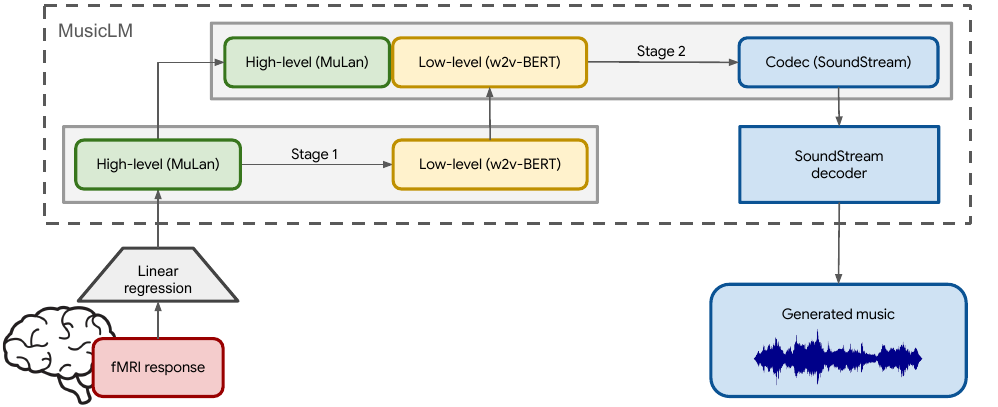}
\end{center}
\caption{Visual representation of {\musiclm} \citep{musiclm} in the context for fMRI decoding. Rounded-rectangle elements denote embeddings/tokens; sharp corners models with parameters. The process begins with an fMRI response, captured from a test subject exposed to music. It is subsequently mapped to a 128-dimensional {\mulan} embedding via linear regression. The first stage of {\musiclm} then refines the {\mulan} embedding into low-level representation of {\wvbert} tokens with temporal information. The subsequent stage, informed by both the output of the previous stage and the {\mulan} embedding, generates tokens for the SoundStream audio codec. In the last step these are transformed into a waveform through a SoundStream decoder.}
\label{fig:musiclm-architecture}
\end{figure}

{\musiclm} \citep{musiclm} is a conditional music generation model. Conditioning signals include -- but are not limited to -- text, other music, and melody.
In our decoding pipeline {\musiclm} is conditioned on a {\mulan} embedding which we compute based on an fMRI response.
Figure~\ref{fig:musiclm-architecture} visualizes the components of {\musiclm}.
Music is generated in two consecutive stages. The first stage learns to map a {\mulan} embedding to a sequence of {\wvbert} tokens.
These tokens used in {\musiclm} are extracted from a {\wvbert} \citep{w2vbert} model's activations in the 7th layer, by clustering them with k-means.

{\musiclm}'s second stage maps the {\wvbert} tokens from the first stage and the {\mulan} embedding to acoustic tokens.
These stem from a SoundStream~\citep{soundstream} model's residual vector quantizer.
The resulting tokens are converted back into audio using a SoundStream decoder.
As in AudioLM~\citep{audiolm}, the second stage is split into a coarse and fine modeling stage.
All three stages are implemented as Transformer models.

\subsection{Decoding: Reconstructing Music from fMRI}
\label{sec:method-decoding}

With \emph{decoding} we refer to attempting the reconstruction of the original stimuli (to which a test subject was exposed) based on their recorded brain activity. This process, shown in Figure~\ref{fig:decoding-pipeline}, can be subdivided into (1)~predicting the music embedding based on fMRI data and (2)~retrieving or generating music based on that embedding.

\paragraph{Music embedding prediction from fMRI data.}

Let $\tR\in\mathbb{R}^{n\times s\times d_\text{fmri}}$ denote the response tensor (obtained via fMRI for each of the five participants), where $n=540$ is the number of stimuli (i.e., 15s music clips), $s=10$ is the number of fMRI scans per stimulus (15s), and $d_\text{fmri}$ is the number of voxels.
$d_\text{fmri}$ varies slightly across subjects depending on the physical size of their brain. For subject~1 it is around 60k.
Our prediction targets are the music embeddings of the stimulus (e.g., {\mulan}), $\tT\in\mathbb{R}^{n\times r\times d_\text{emb}}$, with $r$ being the number of embeddings computed per 15s clip (which depends on the embedding model's window size and the constant step size of 1.5s by which we advance this window). Table~\ref{tab:embedding-types} lists the embeddings, derived from models present in {\musiclm}, that we are considering as candidate \emph{music embeddings} in the Brain2Music architecture from Figure~\ref{fig:decoding-pipeline}.

To align $\tR$ and $\tT$ along the time dimension, we average entries in $\tR$ in a sliding-window fashion to match the time ranges for which feature vectors in $\tT$ were computed.
For example, to predict the {\mulan} embedding ranging from 0s to 10s (due to {\mulan}'s window size being 10s) we rely on the average of five fMRI scans (from 0-1.5s, 1.5-3s, ..., 9s-10.5).
This leaves us with $m:=n\times r$ pairs of response and target, which we split following \cite{nakaidataset}.
We model the relationship between music embeddings and responses with weight matrix $\mW\in\mathbb{R}^{d_\text{fmri} \times d_\text{emb}}$:
\begin{equation}
\label{eq:decoding-lin-reg}
    \hat{\tT}=\tR\mW\,.
\end{equation}
We use an L2-regularized linear regression to estimate $\mW$ on the training dataset.
Note that there is no generalization between different subjects, because of anatomical differences.
For each subject and target dimension, the regression regularization hyperparameter is tuned with five-fold cross validation on the training dataset, while a test split is held out for later evaluation. Details are in Appendix~\ref{sec:appendix-hparam-tuning}.

The training is independently performed for each anatomically defined region of interest (ROI) from the Destrieux atlas \citep{destrieux2010automatic}.
An ROI is a group of voxels.
From all 150~ROIs we select the top $n_\text{ROIs}=6$ with the highest correlation scores (as determined via cross-validation) and create an ensemble model by averaging their predictions.
Concretely that leaves us with ROIs varying in size\footnote{ROI sizes vary between subjects. The top six ROIs of subject 1, for example, which are the most predictive of MuLan embeddings, have the dimensionalities 61, 70, 109, 218, 296, and 706.}. The median across all subjects is 138.5 voxels; the average is 258.6 voxels.
Although the exact location of the ROIs chosen for each subject vary, for all subjects, ROIs are chosen primarily from the auditory cortex.
For a given 15s music stimulus we predict $r$ many embeddings (depending on the chosen embedding type).

\begin{table}[t]
\caption{Different embeddings and their properties: $d$ denotes the embedding dimensionality. $r$ is the number of embeddings we compute per 15s music clip, when we advance the given window with a step size of 1.5s. We also provide the original frequency of the embeddings as provided by the models. After averaging across the given time window for {\soundstream} and {\wvbert} embeddings, the frequency becomes 0.67 Hz for all models, given our step size of 1.5s. For a comparison, we also include properties of the fMRI data of subject 1 (for which $d$ are the dimensions of the top 6 most-correlated ROIs when predicting {\mulanmusic} embeddings).  }
\centering
\small
\label{tab:embedding-types}
\begin{tabular}{lrrrr}
\toprule
\textbf{Embedding} & $d$ & $r$ & \textbf{Original freq. [Hz]} & \textbf{Window size [s]}  \\
\midrule
{\soundstream}-avg & $128$ & $10$  & $50$ & $1.5$ \\
{\wvbert}-avg & $1024$  & $7$ & $25$ & $5$ \\
{\mulanmusic} & $128$ & $4$ & $0.67$ & $10$ \\
{\mulantext} & $128$ & $1$ & - & $15$ \\
\midrule
fMRI (top-6 ROIs) & $1460$ & $10$ & $0.67$ & $1.5$ \\
\bottomrule
\end{tabular}
\end{table}

\paragraph{Music retrieval and music reconstruction.}
We explore two different approaches to derive the original stimulus from the prediction $\hat{\tT}$ namely retrieving similar music from an existing music corpus and generating music with {\musiclm}.

For the retrieval we compute the {\mulan} embeddings for the first 15s of every music clip in the Free Music Archive (FMA) \citep{fma_dataset}.
Unless stated otherwise, we use the \textit{large} variant which contains a wide range of diverse music; concretely, 106,574~music tracks from 161~unbalanced genres.
The \emph{retrieved music} is the audio of the clip whose embeddings are the closest to the predicted one as measured by the cosine similarity.

Alternatively, we can \emph{generate} the music by conditioning the {\musiclm} model (a high-level overview is in Section~\ref{sec:musiclm}) on the predicted embeddings.
The model can then be used to generate music conditionally.
To condition {\musiclm} we average the $r=4$ predicted {\mulan} embeddings along the time dimension.
This is not strictly necessary, but we empirically found the generated outputs to be more stable compared to a version in which we provided all four embeddings to the model.

The two methods, retrieval and generation, have different advantages and disadvantages. The retrieval approach constrains the faithfulness by its limited size.
The predicted embedding could potentially contain rich information about the song which is partially lost by mapping it onto its \textit{nearest} neighbor in the dataset.
The generative model, on the other hand, can in theory generate any kind of music covered by its training distribution, making it conceptually more powerful.
That includes tracks which are not training examples (e.g., combinations of musical concepts). A disadvantage of this method is that the generation model may not adhere precisely to the provided embedding.

\paragraph{Evaluation metrics.} Following the decoding literature ~\citep{image-reconstruction-takagi, park2023sound}, we compute an \emph{identification accuracy} of the predicted $d$-dimensional embeddings with respect to their target embeddings.
Assume there is a matrix of predicted embeddings $\mP\in\mathbb{R}^{n\times d}$ and a matrix (of equal size) containing target embeddings $\mT$. Let $\mC\in\mathbb{R}^{n\times n}$ be computed from $\mP$ and $\mT$, specifically $\mC_{i,j}$ is Pearson correlation coefficient between $i$-th row of $\mP$ and $j$-th row of $\mT$.
The identification accuracy for the $i$-th prediction is defined as:
\begin{equation}
    \text{id acc}_i=
        \frac{1}{n-1}
        \sum_{j=1}^n{\mathbbm{1}\left[C_{i,i} > C_{i,j}\right]}\,.
\end{equation}

The identification accuracy for all examples is simply the average:
\begin{equation}
    \text{id acc}=\frac{1}{n}\sum_{i=1}^n\text{id acc}_i\,.
\end{equation}

This score, ranging from 0 to 1 with 0.5 indicating performance equivalent to random chance, provides a quantified measure of how well an embedding was predicted in relation to other embeddings in the dataset.

An intuitive view on the identification accuracy is that a model with 86\% such accuracy, on average, 14\% of the candidates retrieved will score higher, i.e., have a higher correlation coefficient, than the correct candidate.
In a dataset with 60 examples, the average rank of the correct music clip would be $0.14\times60\approx8$.

As a second metric we also use \emph{top-$n$ class agreement} based on the LEAF~\citep{zeghidour2021leaf} classifier operating on AudioSet classes~\citep{audioset}.
In this context, we compute the per-class probabilities for original and reconstructed music. We then look at three groups of music-related classes, namely genres, instruments, and moods.
For each group we compute the top-$n$ agreement measuring how much overlap there is between the top-$n$ most probable class labels computed for original and reconstruction.
The full list of AudioSet class names in each group is in Appendix~\ref{sec:appendix-audioset-groups}.

\subsection{Encoding: Whole-brain Voxel-wise Modeling}
\label{sec:method-encoding}

To interpret the internal representations of {\musiclm}, we examine the correspondence between them and recorded brain activity.
More specifically, we build whole-brain voxel-wise encoding models to predict fMRI signals using different music embeddings occurring in {\musiclm}: audio-derived embeddings ({\mulanmusic} and {\wvbert}-avg), and text-derived embeddings ({\mulantext}).

We first build encoding models to predict voxel activity from the audio-derived embeddings: {\mulanmusic} and {\wvbert}-avg. We compare how they are represented differently in the human brain.

Next, we build encoding models using audio-derived {\mulanmusic} and text-derived {\mulantext} embeddings to predict fMRI signals. This allows us to explore the differences between these two types of embeddings. The text-derived embeddings are particularly interesting to study, because they can -- by definition -- only represent the high-level information contained in the music caption they are computed from. The {\mulantext} embeddings we use have a 1:1 correspondence to GTZAN clips and are computed by inferring {\mulan}'s text tower (a fine-tuned BERT model) for a given GTZAN clip's text caption. Caption examples are in Section~\ref{sec:appendix-caption-data}.

Finally, we conduct a control analysis to determine whether {\mulanmusic} encompasses more than genre information. To accomplish this, we compare the prediction performance of the {\mulanmusic} model versus one-hot vectors of music genre for each GTZAN clip.

The training data preparation is done in the same manner as in the decoding (outlined in Section~\ref{sec:method-decoding}).
The modeling problem is inverse to Equation~\ref{eq:decoding-lin-reg}, i.e., fMRI responses are predicted based on different embeddings.

Model weights are estimated from training data using L2-regularized linear regression, and subsequently applied to test data. We estimate weights of the model from training data, and regularization parameters are explored during the training using five-fold cross-validation.
For evaluation, we use Pearson's correlation coefficients between predicted and measured fMRI signals.
We compute statistical significance by comparing the estimated correlations to the null distribution of correlations between two independent Gaussian random vectors of the same length of the test data.
We set the threshold for statistical significance at $P<0.05$, and corrected for multiple comparisons using the FDR procedure.
For {\mulantext} and one-hot music genre vectors, we perform up-sampling to match {\mulanmusic}'s sampling rate.

\section{Results}
\label{sec:results}

\subsection{Decoding (fMRI to Music)}
\label{sec:results-dec}

\paragraph{Music embedding prediction.}
Going from fMRI to music requires the prediction of an intermediate music representation, that is, selecting the music embedding to use in Figure~\ref{fig:decoding-pipeline}. In our architecture, the choice of the music embedding represents a bottleneck for the subsequent music generation:
Only if we can predict music embeddings close to music embeddings of the original stimulus heard by the subject, will we be able to generate music that is similar to the original stimulus with {\musiclm}.
The reconstruction quality is further constrained by what the embedding can capture.

Results obtained when predicting different embedding types are reported in Table~\ref{tab:quant-emb-decoding-results}. We find that {\mulanmusic} embeddings can be more accurately predicted from fMRI signals than {\mulantext}, {\wvbert}-avg, or {\soundstream}-avg embeddings.
When predicting {\mulanmusic} embeddings, we observe the highest correlation as well as the highest identification accuracy.
Three reasons may contribute to this: (1)~The {\mulantext} and {\mulanmusic} embeddings, based on their properties mentioned in Table~\ref{tab:embedding-types}, may be more closely aligned to the high level of abstraction represented in the fMRI data than the other embedding types. (2)~The heuristics we applied to align {\wvbert} and {\soundstream} embeddings to the fMRI data may not be optimal and may require further analysis.
(3)~{\mulanmusic} embeddings have access to more musical information than {\mulantext} embeddings because they are computed from the audio signal.

Based on this finding, in the remainder of this section, we use the fMRI data to predict {\mulanmusic} embeddings (for brevity referred to as {\mulan} embeddings) and use them to reconstruct the original stimulus.
However, we include {\wvbert}-avg in the evaluation and the encoding analysis (Section~\ref{sec:method-encoding}) to provide a comparison between two embeddings of different semantic level.

\begin{table}[t]
\caption{Comparison of different music embedding prediction targets (fMRI-to-embedding). The reported error is the standard deviation across five test subjects. We provide metrics on the test dataset, which represent the main results of this table, together with their respective values on the training dataset. Identification accuracy and correlation are computed between embeddings of GTZAN audio (computed with the embedding model listed in the respective row) and the embeddings predicted by the regression model. An analysis of the regression regularization hyperparameter tuning, which is critical to avoid overfitting, can be found in Appendix~\ref{sec:appendix-hparam-tuning}. }
\centering
\small
\label{tab:quant-emb-decoding-results}
\begin{tabular}{lrrrr}
\toprule
\textbf{Embedding}  & \multicolumn{2}{l}{\textbf{Identification accuracy}}               & \multicolumn{2}{l}{\textbf{Correlation}}                 \\
 & \multicolumn{1}{l}{Test} & \multicolumn{1}{l}{Train} & \multicolumn{1}{l}{Test} & \multicolumn{1}{l}{Train} \\ \midrule
{\soundstream}-avg & $0.674\pm0.016$ & $0.764\pm0.029$ & $0.184\pm0.020$ & $0.255\pm0.009$\\
{\wvbert}-avg & $0.837\pm0.005$             & $0.941\pm0.007$             & $0.113\pm0.003$ & $0.167\pm0.006$             \\
{\mulantext} & $0.817\pm0.014$ & $0.877\pm0.009$                        & $0.181\pm0.012$ & $0.245\pm0.009$           \\
{\mulanmusic} & $0.876\pm0.015$ & $0.992\pm0.003$                        & $0.307\pm0.016$ & $0.538\pm0.023$           \\
\bottomrule
\end{tabular}
\end{table}

\paragraph{Qualitative reconstruction results.}

\begin{figure*}[ht]
  \centering
  \makebox[\textwidth][c]{\includegraphics[width=\linewidth]{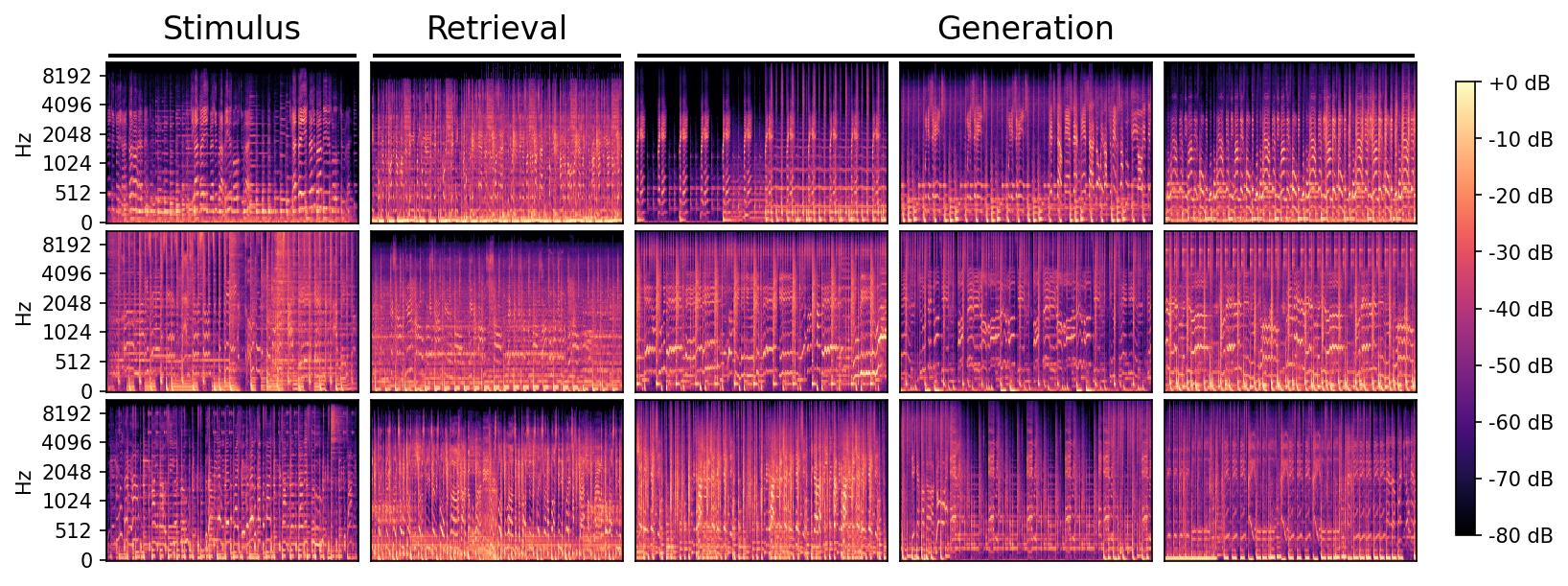}}
   \caption{Spectrograms of different music clips: The left-most column contains the stimulus which subjects were exposed to. To the right is the music retrieved from FMA and three clips sampled from {\musiclm}. Both generation and retrieval are done via {\mulan} embeddings. It is visually perceptible that spectrograms in the same row resemble similarities. Audio examples (randomly sampled, one per genre) can be found at \website{\#ret-vs-gen}}
   \label{fig:retr-vs-gen-qualitative}
\end{figure*}

The two ways of reconstructing music which we compare are retrieval from FMA and generation with {\musiclm}.
Figure~\ref{fig:retr-vs-gen-qualitative} contains qualitative results for subject 1, comparing the original stimulus to music retrieved from FMA based on a predicted {\mulan} embedding and the music clips sampled from {\musiclm}.
Note that sampling multiple clips and examining their differences is one way of qualitatively determining which information {\musiclm} adds to what the {\mulan} embedding contains.
We find that both retrieved and generated constructions are semantically similar to the original stimulus, e.g., with respect to genre, vocal style, overall mood.
The temporal structure of the stimulus is often not preserved in the reconstruction.
There are also failure cases in which the reconstruction is from an entirely different genre.

In Figure~\ref{fig:subject-comparison-qual} we perform a comparison of retrieval and generation across the five different subjects.
The main finding is that qualitatively, the reconstruction is overall of consistent quality across all five subjects.
This is not necessarily a given, when dealing with fMRI data, because of differences in subjective experiences, and suggests our method is robust.

\paragraph{Quantitative reconstruction evaluation.}

\begin{figure*}[t]
  \centering
  \makebox[\textwidth][c]{\includegraphics[width=1.2\linewidth]{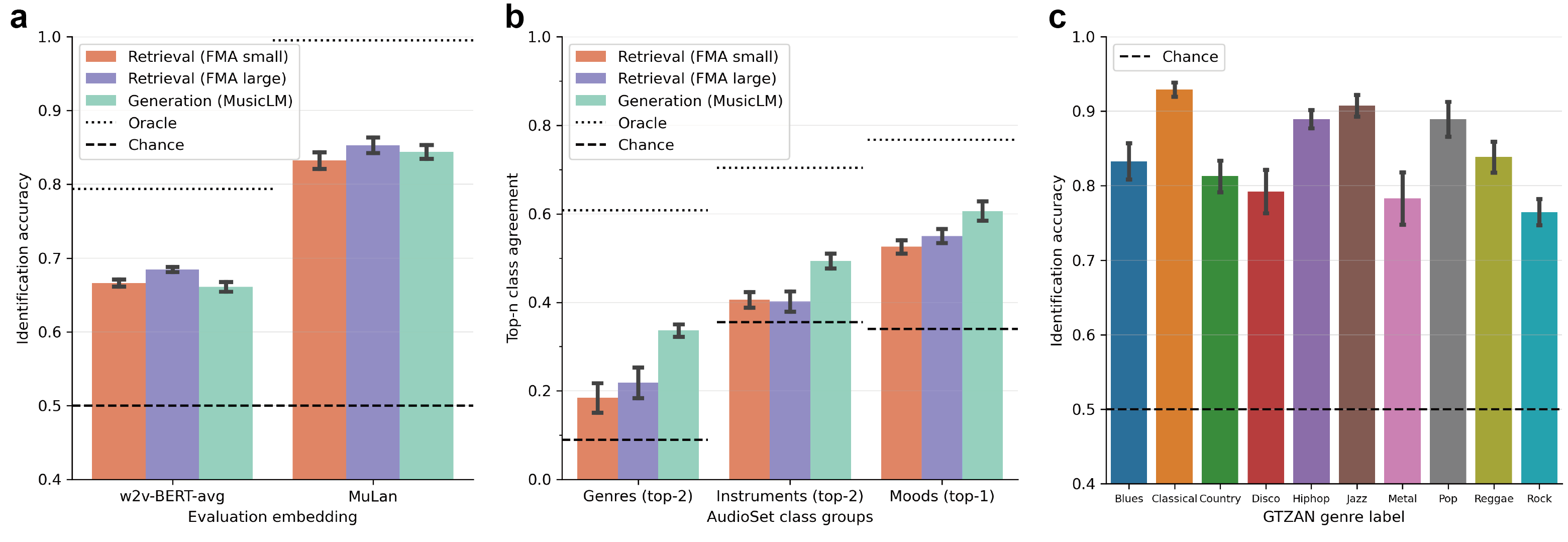}}

   \caption{Main quantitative results of the decoding, i.e., music reconstruction. The dashed, horizontal lines (chance) indicate the performance of a random music predictor (sampling from the FMA dataset). The dotted lines (oracle) are the oracle performance, corresponding to the performance achieved by a regressor which would predict exactly the ground truth {\mulan} embedding of GTZAN. Error bars indicate standard error of the mean across five subjects. \textbf{a} Identification accuracy for different evaluation embeddings. Identification accuracy computed between the embeddings of the original stimulus and the embeddings of the reconstruction. For embedding the music we consider two embedding types. The reconstructed music is more similar to the stimulus it was derived from with respect to high-level embeddings ({\mulan}) than the low-level {\wvbert}-avg. Differences between generation and retrieval on FMA large (about 106k~clips) are marginal, whereas retrieving from FMA small (8k~clips) is overall worse. \textbf{b} AudioSet top-$n$ class agreement for different groups of AudioSet classes. A list of the classes in each group is in Appendix~\ref{sec:appendix-audioset-groups}. Generation is here significantly superior to retrieval from FMA (both small and large). The worst performance -- relative to chance and oracle -- is attained on the instrument agreement. \textbf{c} Identification accuracy (based on {\mulan} embeddings of original and generated music) shown separately for each of the GTZAN genres. The model performance is consistent across all genres.}
   \label{fig:main-quant-results}
\end{figure*}

Figure~\ref{fig:main-quant-results} shows the main quantitative results.
We use two quantitative measures to evaluate the similarity of reconstructed music and original stimulus: identification accuracy (for two embeddings of different semantic level of abstraction) and AudioSet top-$n$ class agreement.

The quality limit imposed by retrieval from FMA is estimated via an \textit{oracle} predictor.
It simply bypasses the linear regression and instead retrieves an FMA clip based on the original music stimulus' {\mulan} embedding.
It simulates the retrieval performance we would achieve if our fMRI to {\mulan} prediction was perfect.
Performance achieved by a model sampling randomly from FMA is indicated by the \textit{chance} result in the plots.

Overall we observe significant above-chance performance on all metrics, establishing strong support for our ability to extract musical information from the fMRI scans.
The identification accuracy comparison across different embedding types hints at our reconstruction to be most faithful to the original stimulus with respect to high-level semantic features as captured by {\mulan}.
While this might seem unsurprising, given {\mulan} is the target embedding of our prediction, it is not necessarily granted that high-level semantic information about perceived music is contained in the recorded brain response in the first place.

Whether or not low-level information is contained in the reconstruction is measured by the identification accuracy on {\wvbert}-avg.
Note, however, that the results are confounded by the embedding averaging we perform for {\wvbert} to align the temporal resolution of the embeddings with that of the fMRI scans.
However, the numbers are in-line with the qualitative observation we make, that is, low-level, acoustic features are relatively not well aligned, whereas the overall music style is.

We further observe consistent prediction performance across different genres (as labeled in the GTZAN dataset).
The highest accuracy is achieved on classical music, which is likely due to its distinctive musical style.

\subsection{Encoding (Brain Activity Prediction)}
\paragraph{Comparison between different audio-derived embeddings.}

\begin{figure*}[t]
  \centering
  \includegraphics[width=1\linewidth]{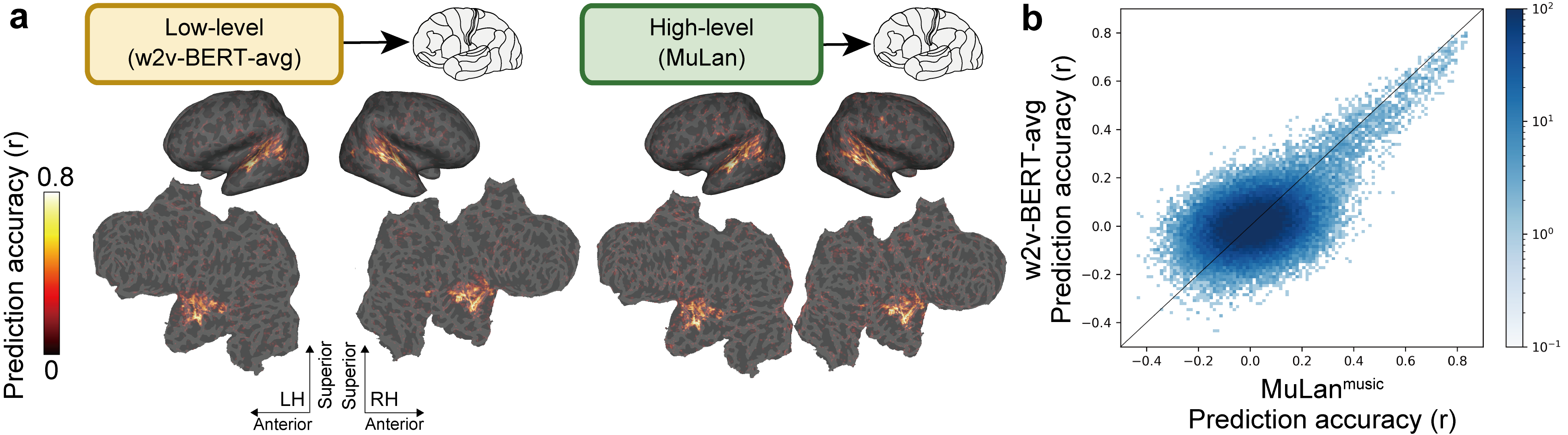}

   \caption{\textbf{a} Prediction performance (measured using Pearson's correlation coefficients) for the voxel-wise encoding model applied to held-out test music on subject 1, projected onto the inflated (top, lateral and medial views) and flattened cortical surface (bottom, occipital areas are at the center), for both left and right hemispheres. Brain regions with significant accuracy are colored (all colored voxels $P<0.05$, FDR corrected). \textbf{b} Density plot of the {\mulanmusic} (x-axis) versus {\wvbert}-avg (y-axis) model prediction accuracy. Darker colors indicate a higher number of voxels in the corresponding bin.}
   \label{fig:encoding-compare-audio-derived-features}
\end{figure*}

\label{sec:results-enc} Figure~\ref{fig:encoding-compare-audio-derived-features}a shows the prediction accuracy of the encoding models for different types of audio-derived embeddings of music within {\musiclm}: {\mulan} and {\wvbert}-avg.
{\mulan} embeddings tend to have higher prediction performance in the lateral prefrontal cortex than {\wvbert}-avg, suggesting that {\mulan} captures high-level music information processed in the human brain. However, when we focus on the auditory cortex, both of the embeddings have some degree of correspondence with human brain activity in the auditory cortex. Given that the text-music model used in this study was not brain-inspired compared to the previous deep learning model such as convolutional neural network, it is intriguing that this correspondence with the brain emerged.
In addition, although each embedding represents different levels of audio-derived embeddings from low ({\wvbert}-avg) to high ({\mulan}), they predicted fairly similar brain regions within the auditory cortex. Figure~\ref{fig:encoding-compare-audio-derived-features}b further confirms that well predicted voxels are largely overlapping between two embeddings. These results suggest that, unlike the visual cortex \citep{image-reconstruction-takagi}, there is not as strong of a hierarchical functional differentiation of audio-derived embeddings in the auditory cortex as previously thought. Note, as we mentioned at \ref{sec:results-dec}, that the results are confounded by the embedding averaging we perform for {\wvbert}. Please see also the limitations in Section~\ref{sec:limitations}. We provide additional results for all subjects in Figure~\ref{supfig:encoding-compare-mulantext-vs-mulanmusic}.

\paragraph{Comparison between audio- and text-derived {\mulan} embeddings.}
In the previous analysis, we confirmed that functional differentiation within the auditory cortex is not as strong with respect to hierarchical audio-derived embeddings.
We next investigate how much text-derived, abstract information about music is differently represented in the auditory cortex compared to audio-derived embeddings.

\begin{figure*}[ht]
  \centering
  \includegraphics[width=1\linewidth]{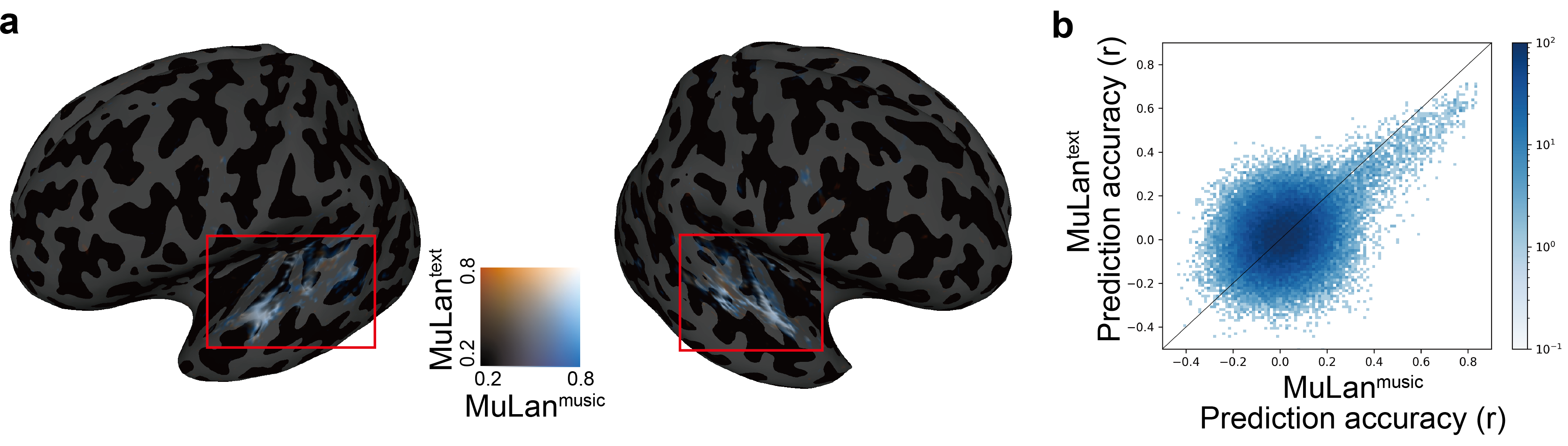}

   \caption{\textbf{a}~Prediction performance on subject 1 for {\mulantext} model versus {\mulanmusic} model. All colored voxels $P<0.05$, FDR corrected. The area in the red rectangle corresponds to the auditory cortex. \textbf{b}~Density plot of the {\mulanmusic} (x-axis) versus {\mulantext} (y-axis) model prediction accuracy.}
   \label{fig:encoding-compare-semantics}
\end{figure*}

Figure~\ref{fig:encoding-compare-semantics}a shows the prediction performance of the encoding models for {\mulantext} versus {\mulanmusic}.
We provide additional results for all subjects in Figure~\ref{supfig:encoding-compare-mulantext-vs-mulanmusic}.
It shows that for some subjects, the inner side of the sulcus represents music stronger than outer side. However, there still seems to be modest functional differentiation in the brain. Although these two representations are trained to match \citep{mulan}, due to the many-to-many nature of text and music pairings the objective cannot be achieved perfectly.
We show that, from a neuroscience perspective, {\mulanmusic} and {\mulantext} actually acquired fairly similar representations. Figure~\ref{fig:encoding-compare-semantics}b further confirms that well-predicted voxels are largely overlapping between two embeddings.

\subsection{Generalization Beyond Music Genre}
\label{sec:past-genre}

\begin{figure*}[t]
  \centering
  \includegraphics[width=1\linewidth]{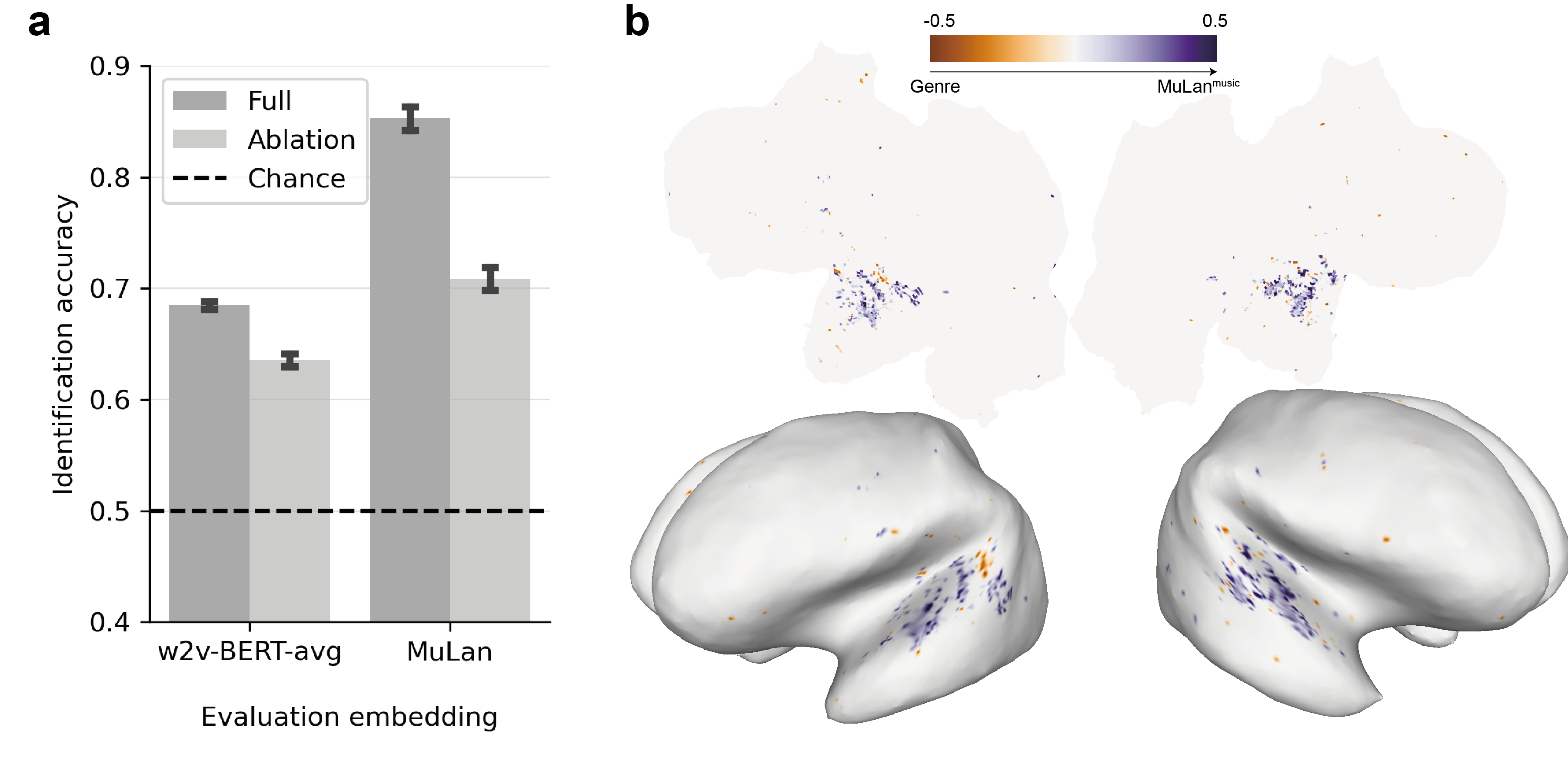}

   \caption{\textbf{a}~Comparison of identification accuracy between the model trained and tested using full data, as shown in Figure~\ref{fig:main-quant-results} (full) and those models trained with one-genre-out ablation data and tested on the ablated genre (ablation). \textbf{b}~Comparing prediction performance obtained by {\mulanmusic} and one-hot genre representation on subject~1. For display purposes, only the voxels where the prediction performance of both models is greater than $0.4$ (almost all of which are within the auditory cortex) are colored.}
   \label{fig:beyond-genre}
\end{figure*}

We next investigate whether our model can generalize to the music genre that was not used during training. To do so, we ablate one genre during training and determine the identification accuracy on clips of the held-out genre (from the test set).
Figure~\ref{fig:beyond-genre}a shows that our model performs significantly above chance on the unseen music genres. This suggests that our reconstruction method is generalizing beyond the genres present in the training data.

We further compare the prediction performances of the {\mulanmusic} and genre models to test whether our encoding model captures information beyond music genres. We find that, compared to the music genres vectors, the {\mulanmusic} embeddings predict activity in the auditory cortex more broadly and with higher accuracy (Figure~\ref{fig:beyond-genre}b).
This is another piece of evidence that our model might predict beyond the music genre information. We provide additional results for all subjects in Figure~\ref{supfig:encoding-beyond-genre}.

\section{Limitations and Future Work}
\label{sec:limitations}

\paragraph{Limitations.} The powerful music generation model we use converts a {\mulan} embedding into music.
Information that is not contained in the embedding, but required to produce a music clip, is added by the model.
In the retrieval case, it is even guaranteed that the reconstruction is musical, because it is directly pulled from a dataset of music.
While this leads to impressive, human-digestible results, it also suggests a higher level of reconstruction quality than there may be.

The main three factors limiting the reconstruction quality are:
\begin{enumerate}
    \item how much information we can extract with linear regression from the fMRI data,
    \item what the chosen music embedding -- in our case {\mulan} -- can capture, and
    \item the limitations of the music retrieval (dataset size and diversity) or generation (generative model capabilities).
\end{enumerate}
In the encoding analysis in Section~\ref{sec:results-enc} we investigate the limitations of (1 + 2) together by predicting brain activity for different embedding types.
We disentangle (2 + 3) from (1) by showing the maximum attainable performance as an oracle dashed-line in Figure~\ref{fig:main-quant-results}. It shows, for example, that top-$1$ class agreement on moods cannot exceed 78\% given the {\mulan} embedding and FMA retrieval dataset choice.
To observe the variability of (3), we experiment with three different reconstruction processes, i.e., two sizes of FMA for retrieval and a generative model.
Further investigation of (1), (2), and (3) remains an open challenge.

The coarse temporal sampling rate of fMRI (1.5s, in the present study) is a limitation of the present study.
However, it is noteworthy that even at the fMRI sampling rate of 2.5s, \cite{Santoro2017} showed temporal specificity at 200ms by using multi-voxel patterns.
Different voxels might have information about different frequency bands, which may collectively contribute to this result.
How much information can actually be retrieved from fMRI is a subject for future research (see \cite{Nishimoto2011} for reconstructing perceived natural movies from fMRI data using frequency-decomposed voxel-wise representations).

The terminology \textit{music reconstruction} may suggest there is a direct reconstruction loss between music stimuli and reconstructed music.
This is -- given the limited amount of data -- clearly not feasible in our setup.
We refer to the overall system as a \textit{music reconstruction}, whereas one might argue that the reconstruction target is actually {\mulan}.
There is currently no clear path to end-to-end music reconstruction with fMRI data.

Lastly, the temporal averaging of {\soundstream} and {\wvbert} embeddings likely worsens their expressiveness.
The role which the embeddings of these models play in {\musiclm}, differs from their application here, in terms of the temporal resolution, which we indicate by adding the suffix ``-avg'' to them whenever the averaged version is used.
Determining which information they contain post-averaging and, orthogonally, evaluating on further measures of music similarity are next steps. This applies to our encoding and decoding studies alike.

\paragraph{Future work.} We perform music reconstruction from fMRI signals that were recorded while human subjects were listening to music stimuli through headphones.
An exciting next step is to attempt the reconstruction of music or musical attributes from a subject's \emph{imagination}.
In such a setting a subject imagines a music clip they know well and the decoding analysis examines how faithfully the imagination can be reconstructed.
This has not been explored yet and would qualify for an actual \emph{mind reading} label.

Comparing the reconstruction properties among a diverse subjects with different musical expertise is another possible direction of exploration.
Comparing, for example, the reconstruction quality between subjects who are professional conductors or musicians and subjects with barely any self-reported music experience could give insights into differences in subjective perception.

\section{Conclusion}
\label{sec:conclusion}

With {\mindmusic} we explored the exciting research direction of reconstructing music from recorded human brain activity.
By conditioning {\musiclm} on a dense music embedding predicted from fMRI data, we were able to generate music which resembles the original music stimuli on a semantic level.

We also investigated the connection between a text-to-music model and the human brain in a quantitative manner by constructing an encoding model.
Specifically, we assessed where and to what extent high-level semantic information and low-level acoustic features of music are represented in the human brain. Although text-to-music models are rapidly developing, their internal processes are still poorly understood. This study is the first to provide a quantitative interpretation from a biological perspective.

Given the nascent stage of music generation models, this work is just a first step. Future work may improve the temporal alignment between reconstruction and stimulus or explore the reconstruction of music from pure imagination.

\section*{Acknowledgements}
We would like to thank Aren Jansen for his review of a draft of this paper. SN was supported by MEXT/JSPS KAKENHI JP18H05522 as well as JST CREST JPMJCR18A5 and ERATO JPMJER1801.

\newpage
\bibliography{bibliography}

\begin{thebibliography}{46}
\providecommand{\natexlab}[1]{#1}
\providecommand{\url}[1]{\texttt{#1}}
\expandafter\ifx\csname urlstyle\endcsname\relax
  \providecommand{\doi}[1]{doi: #1}\else
  \providecommand{\doi}{doi: \begingroup \urlstyle{rm}\Url}\fi

\bibitem[Agostinelli et~al.(2023)Agostinelli, Denk, Borsos, Engel, Verzetti,
  Caillon, Huang, Jansen, Roberts, Tagliasacchi, Sharifi, Zeghidour, and
  Frank]{musiclm}
Andrea Agostinelli, Timo~I. Denk, Zalán Borsos, Jesse Engel, Mauro Verzetti,
  Antoine Caillon, Qingqing Huang, Aren Jansen, Adam Roberts, Marco
  Tagliasacchi, Matt Sharifi, Neil Zeghidour, and Christian Frank.
\newblock Musiclm: Generating music from text, 2023.

\bibitem[Allen et~al.(2018)Allen, Moerel, Lage-Castellanos, {De Martino},
  Formisano, and Oxenham]{ALLEN201860}
Emily~J. Allen, Michelle Moerel, Agustín Lage-Castellanos, Federico {De
  Martino}, Elia Formisano, and Andrew~J. Oxenham.
\newblock Encoding of natural timbre dimensions in human auditory cortex.
\newblock \emph{NeuroImage}, 166:\penalty0 60--70, 2018.
\newblock ISSN 1053-8119.
\newblock \doi{https://doi.org/10.1016/j.neuroimage.2017.10.050}.
\newblock URL
  \url{https://www.sciencedirect.com/science/article/pii/S1053811917308844}.

\bibitem[Alluri et~al.(2012)Alluri, Toiviainen, Jääskeläinen, Glerean, Sams,
  and Brattico]{ALLURI20123677}
Vinoo Alluri, Petri Toiviainen, Iiro~P. Jääskeläinen, Enrico Glerean, Mikko
  Sams, and Elvira Brattico.
\newblock Large-scale brain networks emerge from dynamic processing of musical
  timbre, key and rhythm.
\newblock \emph{NeuroImage}, 59\penalty0 (4):\penalty0 3677--3689, 2012.
\newblock ISSN 1053-8119.
\newblock \doi{https://doi.org/10.1016/j.neuroimage.2011.11.019}.
\newblock URL
  \url{https://www.sciencedirect.com/science/article/pii/S1053811911013000}.

\bibitem[Borsos et~al.(2022)Borsos, Marinier, Vincent, Kharitonov, Pietquin,
  Sharifi, Teboul, Grangier, Tagliasacchi, and Zeghidour]{audiolm}
Zalán Borsos, Raphaël Marinier, Damien Vincent, Eugene Kharitonov, Olivier
  Pietquin, Matt Sharifi, Olivier Teboul, David Grangier, Marco Tagliasacchi,
  and Neil Zeghidour.
\newblock Audiolm: a language modeling approach to audio generation, 2022.

\bibitem[Casey(2017)]{casey2017music}
Michael~A Casey.
\newblock Music of the 7ts: Predicting and decoding multivoxel fmri responses
  with acoustic, schematic, and categorical music features.
\newblock \emph{Frontiers in psychology}, 8:\penalty0 1179, 2017.

\bibitem[Chen et~al.(2023)Chen, Qing, Xiang, Yue, and Zhou]{Chen_2023_CVPR}
Zijiao Chen, Jiaxin Qing, Tiange Xiang, Wan~Lin Yue, and Juan~Helen Zhou.
\newblock Seeing beyond the brain: Masked modeling conditioned diffusion model
  for human vision decoding.
\newblock In \emph{Proceedings of the IEEE/CVF Conference on Computer Vision
  and Pattern Recognition (CVPR)}, 2023.

\bibitem[Chung et~al.(2021)Chung, Zhang, Han, Chiu, Qin, Pang, and Wu]{w2vbert}
Yu-An Chung, Yu~Zhang, Wei Han, Chung-Cheng Chiu, James Qin, Ruoming Pang, and
  Yonghui Wu.
\newblock W2v-bert: Combining contrastive learning and masked language modeling
  for self-supervised speech pre-training, 2021.

\bibitem[Copet et~al.(2023)Copet, Kreuk, Gat, Remez, Kant, Synnaeve, Adi, and
  Défossez]{fbmusicgen}
Jade Copet, Felix Kreuk, Itai Gat, Tal Remez, David Kant, Gabriel Synnaeve,
  Yossi Adi, and Alexandre Défossez.
\newblock Simple and controllable music generation, 2023.

\bibitem[Cross et~al.(2021)Cross, Cockburn, Yue, and O'Doherty]{Cross2021}
Logan Cross, Jeff Cockburn, Yisong Yue, and John~P. O'Doherty.
\newblock Using deep reinforcement learning to reveal how the brain encodes
  abstract state-space representations in high-dimensional environments.
\newblock \emph{Neuron}, 109\penalty0 (4):\penalty0 724--738.e7, February 2021.
\newblock \doi{10.1016/j.neuron.2020.11.021}.
\newblock URL \url{https://doi.org/10.1016/j.neuron.2020.11.021}.

\bibitem[Dale et~al.(1999)Dale, Fischl, and Sereno]{dale1999cortical}
Anders~M Dale, Bruce Fischl, and Martin~I Sereno.
\newblock Cortical surface-based analysis: I. segmentation and surface
  reconstruction.
\newblock \emph{Neuroimage}, 9\penalty0 (2):\penalty0 179--194, 1999.

\bibitem[Defferrard et~al.(2017)Defferrard, Benzi, Vandergheynst, and
  Bresson]{fma_dataset}
Micha\"el Defferrard, Kirell Benzi, Pierre Vandergheynst, and Xavier Bresson.
\newblock {FMA}: A dataset for music analysis.
\newblock In \emph{18th International Society for Music Information Retrieval
  Conference (ISMIR)}, 2017.
\newblock URL \url{https://arxiv.org/abs/1612.01840}.

\bibitem[Destrieux et~al.(2010)Destrieux, Fischl, Dale, and
  Halgren]{destrieux2010automatic}
Christophe Destrieux, Bruce Fischl, Anders Dale, and Eric Halgren.
\newblock Automatic parcellation of human cortical gyri and sulci using
  standard anatomical nomenclature.
\newblock \emph{Neuroimage}, 53\penalty0 (1):\penalty0 1--15, 2010.

\bibitem[Devlin et~al.(2019)Devlin, Chang, Lee, and Toutanova]{devlin2019bert}
Jacob Devlin, Ming-Wei Chang, Kenton Lee, and Kristina Toutanova.
\newblock Bert: Pre-training of deep bidirectional transformers for language
  understanding, 2019.

\bibitem[Dhariwal et~al.(2020)Dhariwal, Jun, Payne, Kim, Radford, and
  Sutskever]{jukebox}
Prafulla Dhariwal, Heewoo Jun, Christine Payne, Jong~Wook Kim, Alec Radford,
  and Ilya Sutskever.
\newblock Jukebox: A generative model for music.
\newblock \emph{arXiv:2005.00341}, 2020.

\bibitem[Donahue et~al.(2023)Donahue, Caillon, Roberts, Manilow, Esling,
  Agostinelli, Verzetti, Simon, Pietquin, Zeghidour, and Engel]{singsong}
Chris Donahue, Antoine Caillon, Adam Roberts, Ethan Manilow, Philippe Esling,
  Andrea Agostinelli, Mauro Verzetti, Ian Simon, Olivier Pietquin, Neil
  Zeghidour, and Jesse Engel.
\newblock Singsong: Generating musical accompaniments from singing, 2023.

\bibitem[Défossez et~al.(2022)Défossez, Copet, Synnaeve, and Adi]{encodec}
Alexandre Défossez, Jade Copet, Gabriel Synnaeve, and Yossi Adi.
\newblock High fidelity neural audio compression, 2022.

\bibitem[Forsgren \& Martiros(2022)Forsgren and Martiros]{riffusion}
Seth Forsgren and Hayk Martiros.
\newblock {Riffusion - Stable diffusion for real-time music generation}, 2022.
\newblock URL \url{https://riffusion.com/about}.

\bibitem[Gemmeke et~al.(2017)Gemmeke, Ellis, Freedman, Jansen, Lawrence, Moore,
  Plakal, and Ritter]{audioset}
Jort~F. Gemmeke, Daniel P.~W. Ellis, Dylan Freedman, Aren Jansen, Wade
  Lawrence, R.~Channing Moore, Manoj Plakal, and Marvin Ritter.
\newblock Audio set: An ontology and human-labeled dataset for audio events.
\newblock In \emph{2017 IEEE International Conference on Acoustics, Speech and
  Signal Processing (ICASSP)}, pp.\  776--780, 2017.
\newblock \doi{10.1109/ICASSP.2017.7952261}.

\bibitem[Ghosal et~al.(2023)Ghosal, Majumder, Mehrish, and
  Poria]{ghosal2023text}
Deepanway Ghosal, Navonil Majumder, Ambuj Mehrish, and Soujanya Poria.
\newblock Text-to-audio generation using instruction-tuned llm and latent
  diffusion model.
\newblock \emph{arXiv preprint arXiv:2304.13731}, 2023.

\bibitem[G{\"u}{\c{c}}l{\"u} \& van Gerven(2015)G{\"u}{\c{c}}l{\"u} and van
  Gerven]{gucclu2015deep}
Umut G{\"u}{\c{c}}l{\"u} and Marcel~AJ van Gerven.
\newblock Deep neural networks reveal a gradient in the complexity of neural
  representations across the ventral stream.
\newblock \emph{Journal of Neuroscience}, 35\penalty0 (27):\penalty0
  10005--10014, 2015.

\bibitem[Hawthorne et~al.(2022)Hawthorne, Jaegle, Cangea, Borgeaud, Nash,
  Malinowski, Dieleman, Vinyals, Botvinick, Simon, et~al.]{perceiverar}
Curtis Hawthorne, Andrew Jaegle, C{\u{a}}t{\u{a}}lina Cangea, Sebastian
  Borgeaud, Charlie Nash, Mateusz Malinowski, Sander Dieleman, Oriol Vinyals,
  Matthew Botvinick, Ian Simon, et~al.
\newblock General-purpose, long-context autoregressive modeling with perceiver
  ar.
\newblock In \emph{International Conference on Machine Learning}, pp.\
  8535--8558. PMLR, 2022.

\bibitem[He et~al.(2015)He, Zhang, Ren, and Sun]{he2015deep}
Kaiming He, Xiangyu Zhang, Shaoqing Ren, and Jian Sun.
\newblock Deep residual learning for image recognition, 2015.

\bibitem[Hoerl \& Kennard(1970)Hoerl and Kennard]{hoerl1970ridge}
Arthur~E Hoerl and Robert~W Kennard.
\newblock Ridge regression: Biased estimation for nonorthogonal problems.
\newblock \emph{Technometrics}, 12\penalty0 (1):\penalty0 55--67, 1970.

\bibitem[Huang et~al.(2022)Huang, Jansen, Lee, Ganti, Li, and Ellis]{mulan}
Qingqing Huang, Aren Jansen, Joonseok Lee, Ravi Ganti, Judith~Yue Li, and
  Daniel P.~W. Ellis.
\newblock Mulan: A joint embedding of music audio and natural language, 2022.

\bibitem[Huang et~al.(2023)Huang, Park, Wang, Denk, Ly, Chen, Zhang, Zhang, Yu,
  Frank, Engel, Le, Chan, Chen, and Han]{noise2music}
Qingqing Huang, Daniel~S. Park, Tao Wang, Timo~I. Denk, Andy Ly, Nanxin Chen,
  Zhengdong Zhang, Zhishuai Zhang, Jiahui Yu, Christian Frank, Jesse Engel,
  Quoc~V. Le, William Chan, Zhifeng Chen, and Wei Han.
\newblock Noise2music: Text-conditioned music generation with diffusion models,
  2023.

\bibitem[Kell et~al.(2018)Kell, Yamins, Shook, Norman-Haignere, and
  McDermott]{Kell2018}
Alexander~J.E. Kell, Daniel~L.K. Yamins, Erica~N. Shook, Sam~V.
  Norman-Haignere, and Josh~H. McDermott.
\newblock A task-optimized neural network replicates human auditory behavior,
  predicts brain responses, and reveals a cortical processing hierarchy.
\newblock \emph{Neuron}, 98\penalty0 (3):\penalty0 630--644.e16, May 2018.
\newblock \doi{10.1016/j.neuron.2018.03.044}.
\newblock URL \url{https://doi.org/10.1016/j.neuron.2018.03.044}.

\bibitem[Koelsch et~al.(2006)Koelsch, Fritz, v.~Cramon, Müller, and
  Friederici]{koelsch2006investigating}
Stefan Koelsch, Thomas Fritz, D.~Yves v.~Cramon, Karsten Müller, and Angela~D.
  Friederici.
\newblock Investigating emotion with music: An fmri study.
\newblock \emph{Human Brain Mapping}, 27\penalty0 (3):\penalty0 239--250, 2006.
\newblock \doi{https://doi.org/10.1002/hbm.20180}.
\newblock URL \url{https://onlinelibrary.wiley.com/doi/abs/10.1002/hbm.20180}.

\bibitem[Lam et~al.(2023{\natexlab{a}})Lam, Tian, Li, Yin, Feng, Tu, Ji, Xia,
  Ma, Song, Chen, Wang, and Wang]{tiktokmusicgen}
Max W.~Y. Lam, Qiao Tian, Tang Li, Zongyu Yin, Siyuan Feng, Ming Tu, Yuliang
  Ji, Rui Xia, Mingbo Ma, Xuchen Song, Jitong Chen, Yuping Wang, and Yuxuan
  Wang.
\newblock Efficient neural music generation, 2023{\natexlab{a}}.

\bibitem[Lam et~al.(2023{\natexlab{b}})Lam, Tian, Li, Yin, Feng, Tu, Ji, Xia,
  Ma, Song, et~al.]{lam2023efficient}
Max~WY Lam, Qiao Tian, Tang Li, Zongyu Yin, Siyuan Feng, Ming Tu, Yuliang Ji,
  Rui Xia, Mingbo Ma, Xuchen Song, et~al.
\newblock Efficient neural music generation.
\newblock \emph{arXiv preprint arXiv:2305.15719}, 2023{\natexlab{b}}.

\bibitem[Liu et~al.(2023)Liu, Chen, Yuan, Mei, Liu, Mandic, Wang, and
  Plumbley]{liu2023audioldm}
Haohe Liu, Zehua Chen, Yi~Yuan, Xinhao Mei, Xubo Liu, Danilo Mandic, Wenwu
  Wang, and Mark~D Plumbley.
\newblock Audioldm: Text-to-audio generation with latent diffusion models.
\newblock \emph{arXiv preprint arXiv:2301.12503}, 2023.

\bibitem[Moeller et~al.(2010)Moeller, Yacoub, Olman, Auerbach, Strupp, Harel,
  and U{\u{g}}urbil]{moeller2010multiband}
Steen Moeller, Essa Yacoub, Cheryl~A Olman, Edward Auerbach, John Strupp, Noam
  Harel, and K{\^a}mil U{\u{g}}urbil.
\newblock Multiband multislice ge-epi at 7 tesla, with 16-fold acceleration
  using partial parallel imaging with application to high spatial and temporal
  whole-brain fmri.
\newblock \emph{Magnetic resonance in medicine}, 63\penalty0 (5):\penalty0
  1144--1153, 2010.

\bibitem[Nakai et~al.(2021)Nakai, Koide-Majima, and Nishimoto]{nakai2021music}
Tomoya Nakai, Naoko Koide-Majima, and Shinji Nishimoto.
\newblock Correspondence of categorical and feature-based representations of
  music in the human brain.
\newblock \emph{Brain and Behavior}, 11\penalty0 (1):\penalty0 e01936, 2021.
\newblock \doi{https://doi.org/10.1002/brb3.1936}.
\newblock URL \url{https://onlinelibrary.wiley.com/doi/abs/10.1002/brb3.1936}.

\bibitem[Nakai et~al.(2022)Nakai, Koide-Majima, and Nishimoto]{nakaidataset}
Tomoya Nakai, Naoko Koide-Majima, and Shinji Nishimoto.
\newblock Music genre neuroimaging dataset.
\newblock \emph{Data in Brief}, 40:\penalty0 107675, 2022.
\newblock ISSN 2352-3409.
\newblock \doi{https://doi.org/10.1016/j.dib.2021.107675}.
\newblock URL
  \url{https://www.sciencedirect.com/science/article/pii/S2352340921009501}.

\bibitem[Naselaris et~al.(2011)Naselaris, Kay, Nishimoto, and
  Gallant]{naselaris2011}
Thomas Naselaris, Kendrick Kay, Shinji Nishimoto, and Jack Gallant.
\newblock Encoding and decoding in fmri.
\newblock \emph{NeuroImage}, 56:\penalty0 400--10, 05 2011.
\newblock \doi{10.1016/j.neuroimage.2010.07.073}.

\bibitem[Nishimoto et~al.(2011)Nishimoto, Vu, Naselaris, Benjamini, Yu, and
  Gallant]{Nishimoto2011}
Shinji Nishimoto, An~T. Vu, Thomas Naselaris, Yuval Benjamini, Bin Yu, and
  Jack~L. Gallant.
\newblock Reconstructing visual experiences from brain activity evoked by
  natural movies.
\newblock \emph{Current Biology}, 21\penalty0 (19):\penalty0 1641--1646,
  October 2011.
\newblock \doi{10.1016/j.cub.2011.08.031}.
\newblock URL \url{https://doi.org/10.1016/j.cub.2011.08.031}.

\bibitem[Norman-Haignere et~al.(2015)Norman-Haignere, Kanwisher, and
  McDermott]{norman2015distinct}
Sam Norman-Haignere, Nancy~G Kanwisher, and Josh~H McDermott.
\newblock Distinct cortical pathways for music and speech revealed by
  hypothesis-free voxel decomposition.
\newblock \emph{neuron}, 88\penalty0 (6):\penalty0 1281--1296, 2015.

\bibitem[Park et~al.(2023)Park, Tsukamoto, Tanaka, and Kamitani]{park2023sound}
Jong-Yun Park, Mitsuaki Tsukamoto, Misato Tanaka, and Yukiyasu Kamitani.
\newblock Sound reconstruction from human brain activity via a generative model
  with brain-like auditory features, 2023.

\bibitem[Santoro et~al.(2017)Santoro, Moerel, Martino, Valente, Ugurbil,
  Yacoub, and Formisano]{Santoro2017}
Roberta Santoro, Michelle Moerel, Federico~De Martino, Giancarlo Valente, Kamil
  Ugurbil, Essa Yacoub, and Elia Formisano.
\newblock Reconstructing the spectrotemporal modulations of real-life sounds
  from {fMRI} response patterns.
\newblock \emph{Proceedings of the National Academy of Sciences}, 114\penalty0
  (18):\penalty0 4799--4804, April 2017.
\newblock \doi{10.1073/pnas.1617622114}.
\newblock URL \url{https://doi.org/10.1073/pnas.1617622114}.

\bibitem[Shen et~al.(2019)Shen, Horikawa, Majima, and Kamitani]{Shen2019}
Guohua Shen, Tomoyasu Horikawa, Kei Majima, and Yukiyasu Kamitani.
\newblock Deep image reconstruction from human brain activity.
\newblock \emph{{PLOS} Computational Biology}, 15\penalty0 (1):\penalty0
  e1006633, January 2019.
\newblock \doi{10.1371/journal.pcbi.1006633}.
\newblock URL \url{https://doi.org/10.1371/journal.pcbi.1006633}.

\bibitem[Takagi \& Nishimoto(2022)Takagi and
  Nishimoto]{image-reconstruction-takagi}
Yu~Takagi and Shinji Nishimoto.
\newblock High-resolution image reconstruction with latent diffusion models
  from human brain activity.
\newblock \emph{bioRxiv}, 2022.
\newblock \doi{10.1101/2022.11.18.517004}.
\newblock URL
  \url{https://www.biorxiv.org/content/early/2022/11/21/2022.11.18.517004}.

\bibitem[Takagi \& Nishimoto(2023)Takagi and Nishimoto]{Takagi2023TechRep}
Yu~Takagi and Shinji Nishimoto.
\newblock Improving visual image reconstruction from human brain activity using
  latent diffusion models via multiple decoded inputs, 2023.
\newblock URL \url{https://arxiv.org/abs/2306.11536}.

\bibitem[Toiviainen et~al.(2014)Toiviainen, Alluri, Brattico, Wallentin, and
  Vuust]{TOIVIAINEN2014170}
Petri Toiviainen, Vinoo Alluri, Elvira Brattico, Mikkel Wallentin, and Peter
  Vuust.
\newblock Capturing the musical brain with lasso: Dynamic decoding of musical
  features from fmri data.
\newblock \emph{NeuroImage}, 88:\penalty0 170--180, 2014.
\newblock ISSN 1053-8119.
\newblock \doi{https://doi.org/10.1016/j.neuroimage.2013.11.017}.
\newblock URL
  \url{https://www.sciencedirect.com/science/article/pii/S1053811913011099}.

\bibitem[Tzanetakis \& Cook(2002)Tzanetakis and Cook]{gtzan}
G.~Tzanetakis and P.~Cook.
\newblock Musical genre classification of audio signals.
\newblock \emph{IEEE Transactions on Speech and Audio Processing}, 10\penalty0
  (5):\penalty0 293--302, 2002.
\newblock \doi{10.1109/TSA.2002.800560}.

\bibitem[Yamins et~al.(2014)Yamins, Hong, Cadieu, Solomon, Seibert, and
  DiCarlo]{yamins2014performance}
Daniel~LK Yamins, Ha~Hong, Charles~F Cadieu, Ethan~A Solomon, Darren Seibert,
  and James~J DiCarlo.
\newblock Performance-optimized hierarchical models predict neural responses in
  higher visual cortex.
\newblock \emph{Proceedings of the national academy of sciences}, 111\penalty0
  (23):\penalty0 8619--8624, 2014.

\bibitem[Zeghidour et~al.(2021{\natexlab{a}})Zeghidour, Luebs, Omran, Skoglund,
  and Tagliasacchi]{soundstream}
Neil Zeghidour, Alejandro Luebs, Ahmed Omran, Jan Skoglund, and Marco
  Tagliasacchi.
\newblock Soundstream: An end-to-end neural audio codec, 2021{\natexlab{a}}.

\bibitem[Zeghidour et~al.(2021{\natexlab{b}})Zeghidour, Teboul,
  de~Chaumont~Quitry, and Tagliasacchi]{zeghidour2021leaf}
Neil Zeghidour, Olivier Teboul, Félix de~Chaumont~Quitry, and Marco
  Tagliasacchi.
\newblock Leaf: A learnable frontend for audio classification,
  2021{\natexlab{b}}.

\end{thebibliography}
\bibliographystyle{bib-style}
\appendix
\setcounter{figure}{0}
\renewcommand\thefigure{\Alph{section}.\arabic{figure}}
\renewcommand\thetable{\Alph{section}.\arabic{table}}

\section{Appendix}
\label{sec:appendix}

\subsection{Method Details}
\label{sec:appendix-method-details}

\subsubsection{fMRI Data Preprocessing}
\label{sec:appendix-fMRI-preprocess}
We pre-process the \textit{music genre neuroimaging dataset}\footnote{Download link: \href{https://openneuro.org/datasets/ds003720/versions/1.0.0}{openneuro.org/datasets/ds003720}} from \cite{nakaidataset} in a same fashion as \cite{nakai2021music}, recited below.
Motion correction is performed for each run using the Statistical Parametric Mapping toolbox (SPM8; Wellcome Trust Centre for Neuroimaging, London, UK; \href{http://www.fil.ion.ucl.ac.uk/spm/}{fil.ion.ucl.ac.uk/spm}).
All volumes are aligned to the first EPI image for each participant.
Low-frequency drift is removed using a median filter with a 240s window.
To augment model fitting accuracy, the response for each voxel is normalized by subtracting the mean response and then scaling it to the unit variance.
We use FreeSurfer \citep{dale1999cortical} to identify the cortical surfaces from the anatomical data and register them with the voxels of the functional data.
We use only cortical voxels as targets of the analysis for each participant. For each participant, we use the voxels identified in the cerebral cortex in the analysis (53,421 to 64,700 voxels per participant).

\subsubsection{Hyperparameter Tuning}
\label{sec:appendix-hparam-tuning}

We use a ridge regression to estimate our model parameters. Citing the \textit{himalaya} documentation\footnote{\href{https://gallantlab.org/himalaya/models.html\#ridge}{gallantlab.org/himalaya/models.html\#ridge}}:
Let $\mX\in\mathbb{R}^{n\times p}$ be a feature matrix with $n$ samples and $p$ features, $\vy\in\mathbb{R}^n$ a target vector, and $\alpha>0$ a fixed regularization hyperparameter. Ridge regression \citep{hoerl1970ridge} defines the weight vector $\vb^*\in\mathbb{R}^p$ as:
\begin{equation}
    \vb^*=\operatorname{arg}\operatorname{min}_{\vb}\lVert\mX\vb-\vy\rVert^2_2+\alpha\lVert\vb\rVert^2_2\,.
\end{equation}
The equation has a closed-form solution $\vb^*=\mM\vy$, where $\mM =
(\mX^\top \mX + \alpha \mI_p)^{-1}\mX^\top \in  \mathbb{R}^{p \times n}$.

\begin{figure*}[ht]
  \centering
  \makebox[\textwidth][c]{\includegraphics[width=1\linewidth]{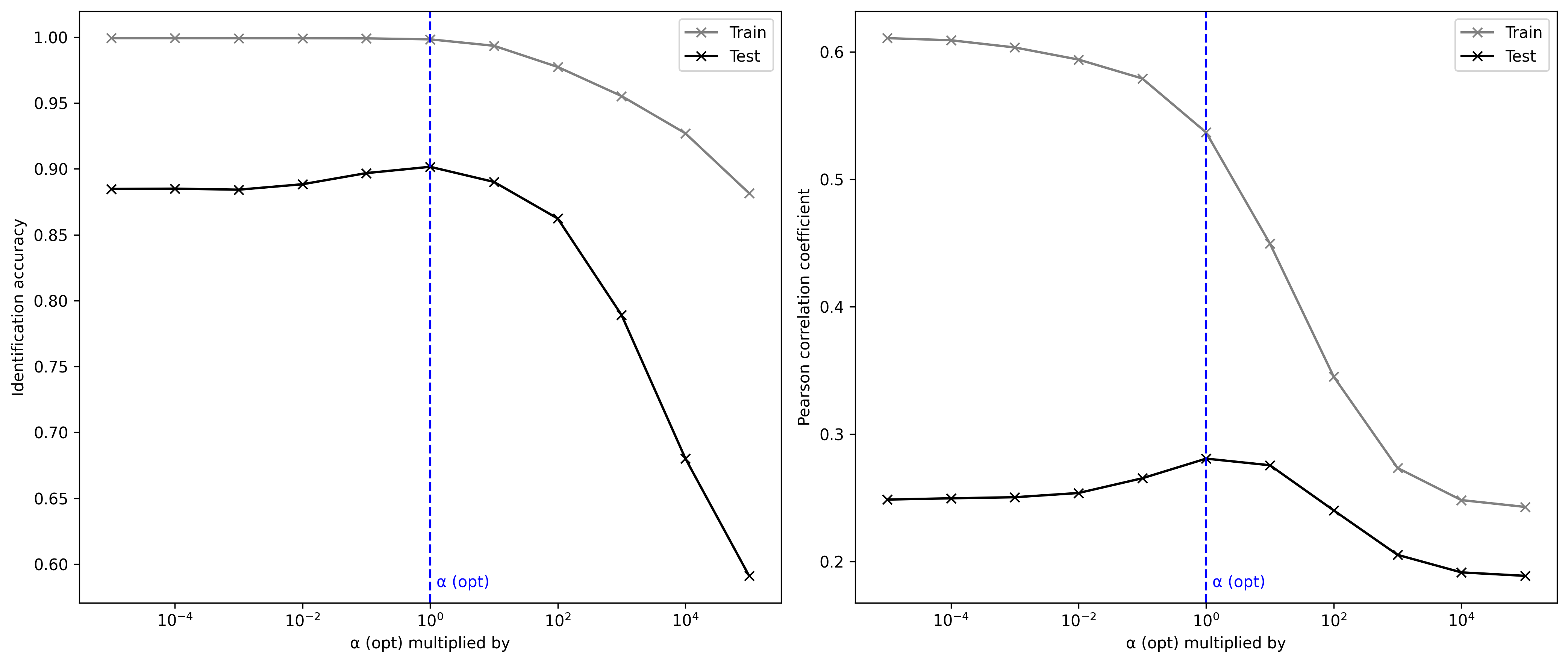}}
  \caption{Performance of the regressor when trained with $\alpha$ values in the neighborhood of the $\alpha$ that was determined to be optimal on the \emph{training split} via cross-validation. The model starts to overfit with lower values of $\alpha$ (to the left) and underfits in the opposite direction. The plot is computed for a model predicting {\mulan} embeddings for subject 1.}
  \label{fig:alpha-hparam}
\end{figure*}

To determine $\alpha$ we run 5-fold cross-validation on the training data. Note that there is one $\alpha$ parameter per regression target, i.e., 128 in our case when predicting {\mulan} embeddings.
We inspect the performance on training and evaluation data around the chosen $\alpha$ vector ``$\alpha$ (opt)'' in Figure~\ref{fig:alpha-hparam}.

\pagebreak
\subsubsection{AudioSet Class Groups}
\label{sec:appendix-audioset-groups}

In our AudioSet evaluation metric we compute the overlap of the top-$n$ most likely classes (between reconstructed and original music) in three different groups. The groups and the $n$ choice are genres (top-2), instruments (top-2), and moods (top-1).

Below is a list of the AudioSet class names of each group:

\small{Genres (top-2): \textit{Pop music, Hip hop music, Rock music, Rhythm and blues, Soul music, Reggae, Country, Funk, Folk music, Middle Eastern music, Jazz, Disco, Classical music, Electronic music, Music of Latin America, Blues, Music for children, New-age music, Vocal music, Music of Africa, Christian music, Music of Asia, Ska, Traditional music, Independent music}}

\small{Instruments (top-2): \textit{Plucked string instrument, Keyboard (musical), Percussion, Orchestra, Brass instrument, Bowed string instrument, Wind instrument, woodwind instrument, Harp, Choir, Bell, Harmonica, Accordion, Bagpipes, Didgeridoo, Shofar, Theremin, Singing bowl, Scratching (performance technique)}}

\small{Moods (top-1): \textit{Happy music, Sad music, Tender music, Exciting music, Angry music, Scary music}}

\FloatBarrier
\subsection{Additional Results of the Decoding}
\label{sec:appendix-reconstruction}

\begin{figure*}[ht]
  \centering
  \makebox[\textwidth][c]{\includegraphics[width=1.25\linewidth]{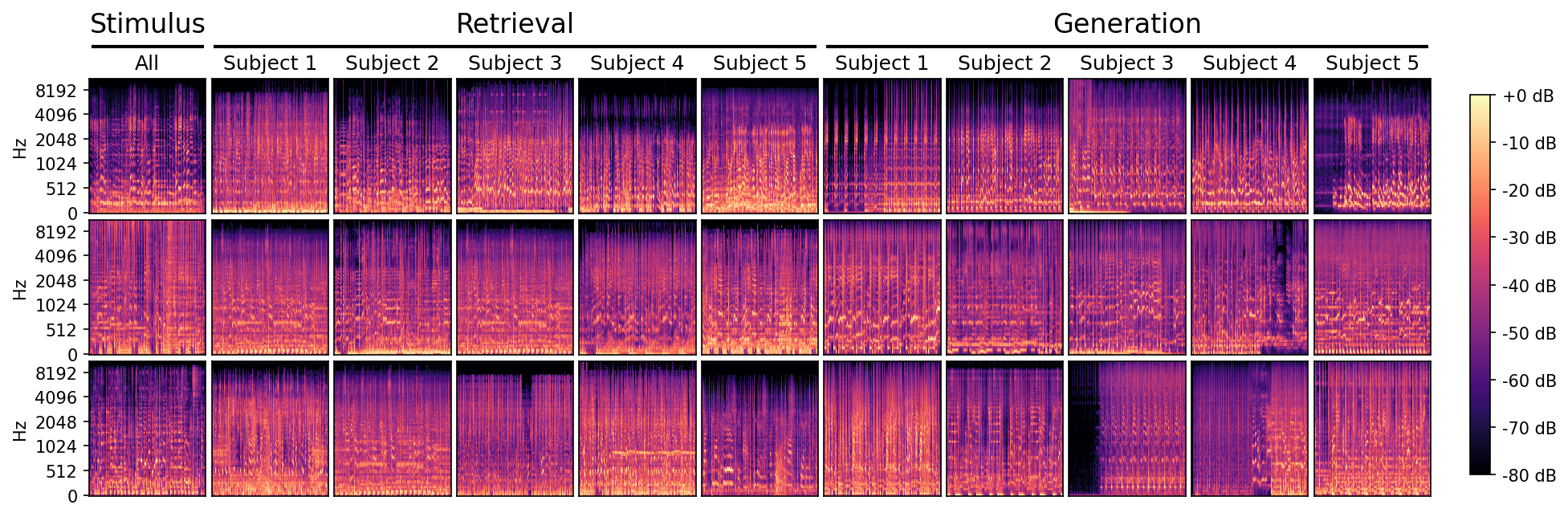}}
  \caption{The comparison shows the spectrograms of music retrieved or generated for different test subjects, who were all exposed to the same stimuli. The predicted embedding is {\mulan}. Audio examples (randomly sampled, one per genre) can be found at \website{\#all-subjects}}
  \label{fig:subject-comparison-qual}
\end{figure*}

\newpage
\FloatBarrier
\subsection{Additional Results of Encoding}
\begin{figure*}[ht]
  \centering
  \includegraphics[width=1\linewidth]{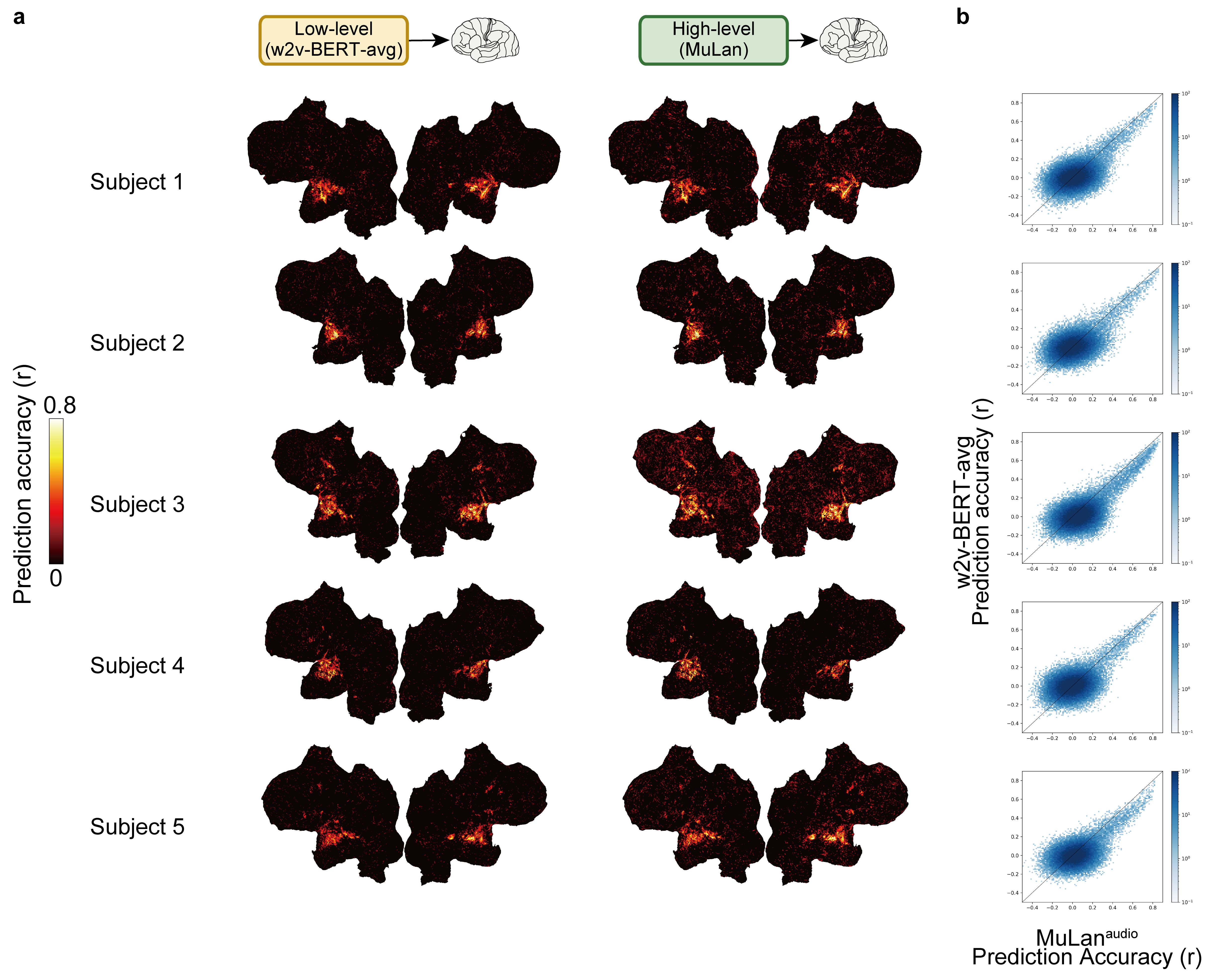}

  \caption{\textbf{a} All subject results for comparing prediction performances between different audio-derived embeddings: {\mulanmusic} and {\wvbert}-avg. \textbf{b} Density plot of the {\mulanmusic} (x-axis) versus {\wvbert}-avg (y-axis) model prediction accuracy. Darker colors indicate a higher number of voxels in the corresponding bin.}
  \label{supfig:encoding-compare-music-derived-features}
\end{figure*}

\begin{figure*}[ht]
  \centering
  \includegraphics[width=1\linewidth]{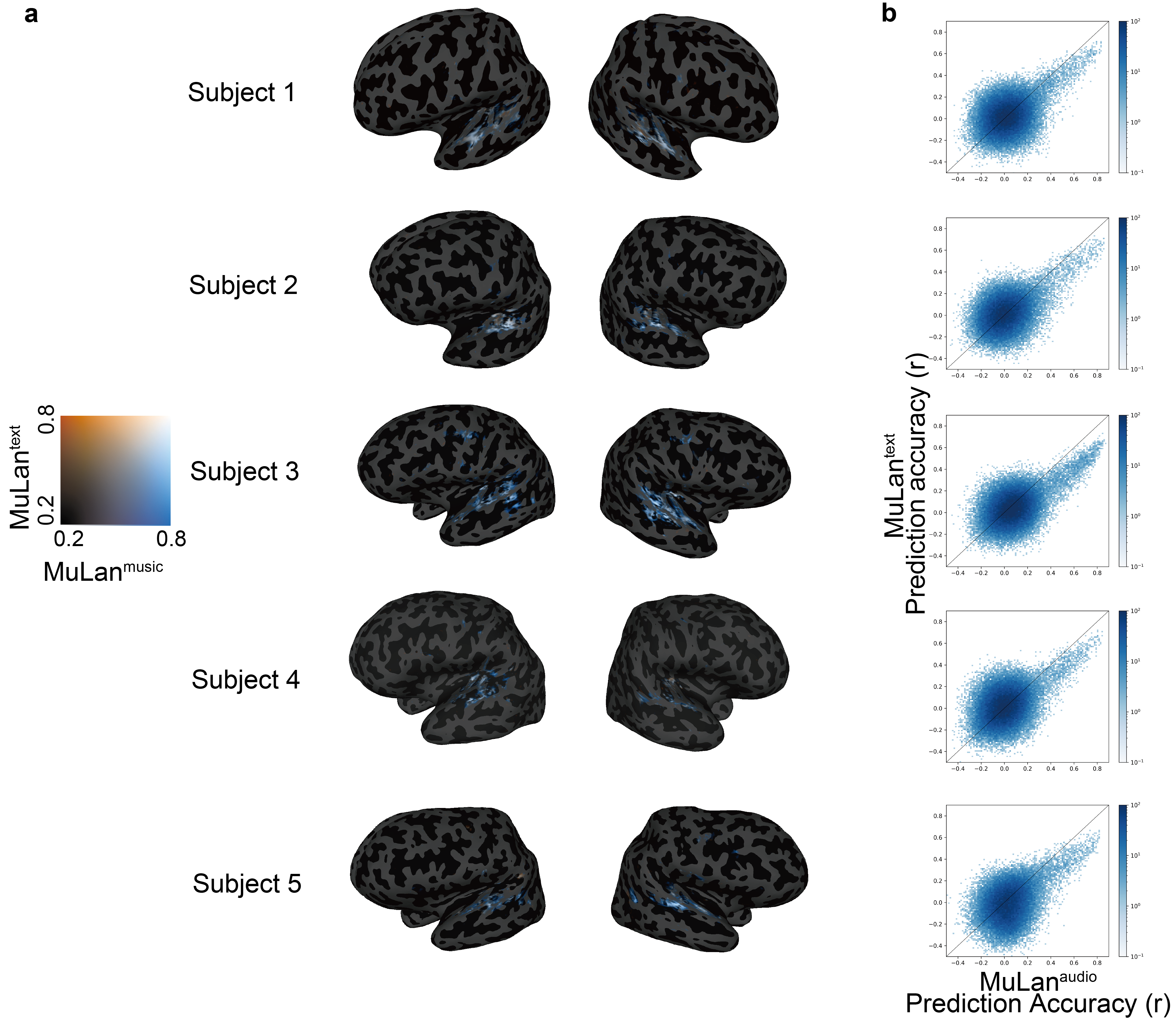}

  \caption{\textbf{a} All subject results for comparing prediction performances between {\mulanmusic} and {\mulantext}. \textbf{b} Density plot of the {\mulanmusic} (x-axis) versus {\mulantext} (y-axis) model prediction accuracy.}
  \label{supfig:encoding-compare-mulantext-vs-mulanmusic}
\end{figure*}

\begin{figure*}[p]
  \centering
  \includegraphics[width=1\linewidth]{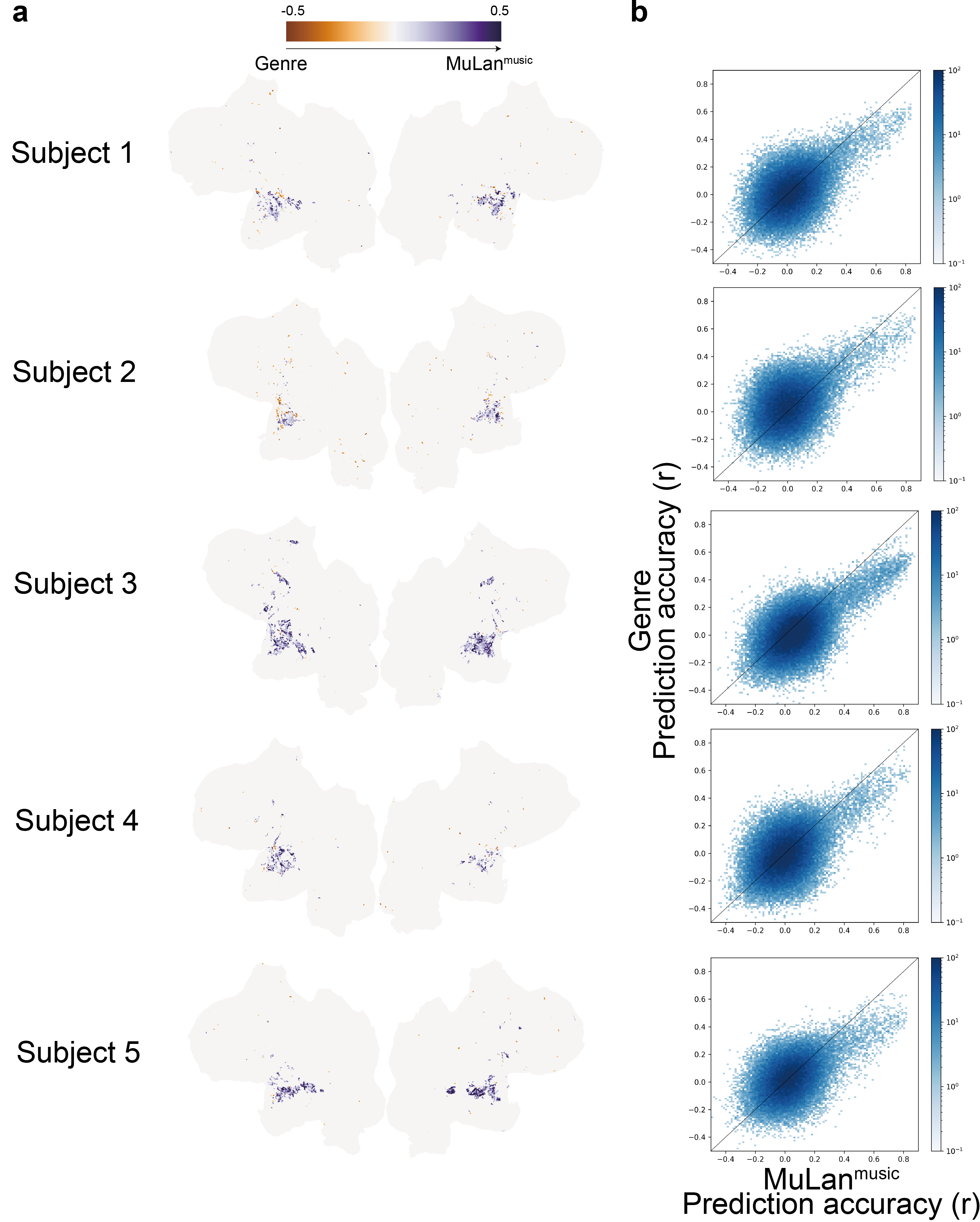}

  \caption{\textbf{a} All subject results for comparing prediction performances between {\mulanmusic} and GTZAN genre label. \textbf{b} Density plot of the {\mulanmusic} (x-axis) versus genre (y-axis) model prediction accuracy. Many voxels are better predicted by {\mulanmusic} than genre.}
  \label{supfig:encoding-beyond-genre}
\end{figure*}

We present additional results for all subjects in Figures \ref{supfig:encoding-compare-music-derived-features}, \ref{supfig:encoding-compare-mulantext-vs-mulanmusic}, and \ref{supfig:encoding-beyond-genre} for Figures \ref{fig:encoding-compare-audio-derived-features}, \ref{fig:encoding-compare-semantics}, and \ref{fig:beyond-genre}.
They show that our results are robust across subjects.

\newpage
\FloatBarrier
\subsection{Text Caption Dataset}
\label{sec:appendix-caption-data}

We release a text caption dataset\footnote{\href{https://www.kaggle.com/datasets/nishimotolab/music-caption-brain2music}{kaggle.com/datasets/nishimotolab/music-caption-brain2music}} for the 540~GTZAN music clips (15s crop) for which fMRI data as recorded.
The captions were collected with the online freelance platform \href{https://coconala.com/}{coconala.com} by human raters, all of which are music professionals (musician, teacher, composer).
The instruction given to the raters is:

\textit{\small{We have numerous 15-second clips for which we’d like you to provide a written description of about four sentences in Japanese (or English). By written description, we mean something that includes an explanation or impression of the music piece, as demonstrated in the following example: ``This is a Drum \& Bass track. It features high-speed scratching from a turntable and includes sampled screams. You can hear a sinister tune being played by the synthesizer. The rhythmic backdrop is composed of fast electronic drum beats. This track seems like it could be used as a soundtrack for a car racing game.''}}

Below are ten example captions; one per genre. A table with audio and captions side by side is at \website{\#gtzan-caps}

\small{blues.00017: \textit{It is lazy blues with a laid-back tempo and relaxed atmosphere. The band structure is simple, with the background rhythm punctuated by bass and guitar cutting. The impressive phrasing of the lead guitar gives the piece a nostalgic impression.}}

\small{classical.00008: \textit{Several violins play the melody. The melody is simple and almost unison, but it moves between minor and major keys and changes expression from one to the other.}}

\small{country.00012: \textit{This is a classic country song. You can hear clear singing and crisp acoustic guitar cutting. The wood bass provides a solid groove with a two-beat rhythm. This is country music at its best. Ideal for nature scenes and homely atmospheres.}}

\small{disco.00004: \textit{This music piece has a disco sound. Vocals and chorus create extended harmonies. The synthesiser creates catchy melodies, while the drumming beats rhythmically. Effective tambourine sounds accentuate the rhythms and add further dynamism. This music is perfect for dance parties, club floors and other scenes of dancing and fun.}}

\small{hiphop.00014: \textit{This is a rap-rock piece with a lot of energy. The distorted guitars are impressive and provide an energetic sound. The bass is an eight beat, creating a dynamic groove. The drums provide the backbone of the rhythm section with their powerful hi-hats. The vocal and chorus interaction conveys tension and passion and draws the audience in.}}

\small{jazz.00040: \textit{This is medium-tempo old jazz with female vocals. The band is a small band similar to a Dixie Jazz formation, including clarinet, trumpet and trombone. The vocal harmonies are supported by a piano and brass ensemble on a four beat with drums and bass.}}

\small{metal.00026: \textit{This is a metal instrumental piece with technical guitar solos and distortion effects. The heavy, powerful bass creates a sense of speed, and the snare, bass and guitar create a sense of unity in unison at the end. It is full of over-the-top playing techniques and intense energy.}}

\small{pop.00032: \textit{Passionate pops piece with clear sound and female vocals. The synth accompaniment spreads out pleasantly and the tight bass grooves along. The beat-oriented drums drive the rhythm, creating a strong and lively feeling. Can be used as background music in cafés and lounges to create a relaxed atmosphere.}}

\small{reggae.00013: \textit{This reggae piece combines smooth, melodic vocals with a clear, high-pitched chorus. The bass is swingy and supports the rhythm, while whistles and samplers of life sounds can be heard. It is perfect for relaxing situations, such as reading in a laid-back café or strolling around town.}}

\small{rock.00032: \textit{This rock piece is characterised by its extended vocals. The guitar plays scenically, while the bass enhances the melody with rhythmic fills. The drums add dynamic rhythms to the whole piece. This music is ideal for scenes with a sense of expansiveness and freedom, such as mountainous terrain with spectacular natural scenery or driving scenes on the open road.}}

\end{document}